\documentclass[aps,prd,preprint,superscriptaddress,amsmath,amssymb,nofootinbib]{revtex4-1}
\usepackage[utf8]{inputenc}
\usepackage[T1]{fontenc}
\usepackage{braket}
\usepackage{hyperref}
\usepackage{graphicx}
\usepackage{subcaption}
\usepackage{xcolor}

\graphicspath{ {./plots/} }

\newcommand{\G}[1]{\Gamma\left(#1\right)}
\newcommand{\ac}{ \alpha_{\mathbb S} }

\begin{document}
	
	\title{Exact Correlation Functions in the Brownian Loop Soup}

	\author{Federico Camia}
	\affiliation{New York University Abu Dhabi, Saadiyat Island, UAE}
	\affiliation{Department of Mathematics, VU University, De Boelelaan 1081a, 1081 HV Amsterdam, The Netherlands}
	\author{Valentino F.\ Foit}
    \affiliation{Center for Cosmology and Particle Physics, New York University, 726 Broadway, New York, NY 10003, USA}
	\author{Alberto Gandolfi}
	\affiliation{New York University Abu Dhabi, Saadiyat Island, UAE}
	\author{Matthew Kleban}
	\affiliation{Center for Cosmology and Particle Physics, New York University, 726 Broadway, New York, NY 10003, USA}
	
	\date{\today}
	
\begin{abstract}
We  compute analytically and in closed form the four-point correlation function in the plane, and the two-point correlation function in the upper half-plane, of layering vertex operators in the two dimensional conformally invariant system known as the Brownian Loop Soup.  These correlation functions depend on multiple continuous parameters: the insertion points of the operators, the intensity of the soup, and the charges of the operators.  In the case of the four-point function there is non-trivial dependence on five continuous parameters: the cross-ratio, the intensity, and three real charges.  The four-point function is crossing symmetric.  We analyze its conformal block expansion and discover a previously unknown set of new conformal primary operators.  

\end{abstract}

\maketitle

\section{Introduction}

The Brownian Loop Soup (BLS) \cite{lawler2003brownian} is a conformally invariant system consisting of closed, self-intersection loops randomly distributed in the plane according to a  conformally invariant measure $\mu^{\text{loop}}$. The measure has one free parameter, an overall normalization $\lambda > 0$, that is called the intensity of the soup and determines the density of the loops.

In earlier work \cite{camia2016brownian}, three of us established that the soup has a central charge $c = 2 \lambda$. Since $c$ is continuous and can be less than one, this demonstrates that the model cannot be unitary for all values of $\lambda$ (the minimal models are the only unitary conformal field theories with $c<1$, and they come in a discrete series). Nevertheless, we were able to identify a set of conformal primary operators with positive conformal dimensions. These operators are exponentials of the form $e^{i \beta N(z)}$, where $\beta$ is a real number (the ``charge'') and $N(z)$ is an integer-valued operator that counts some characteristics of the loops. 

\begin{figure}[t]
    \centering
    \begin{subfigure}[b]{0.45\textwidth}
        \includegraphics[width=\textwidth]{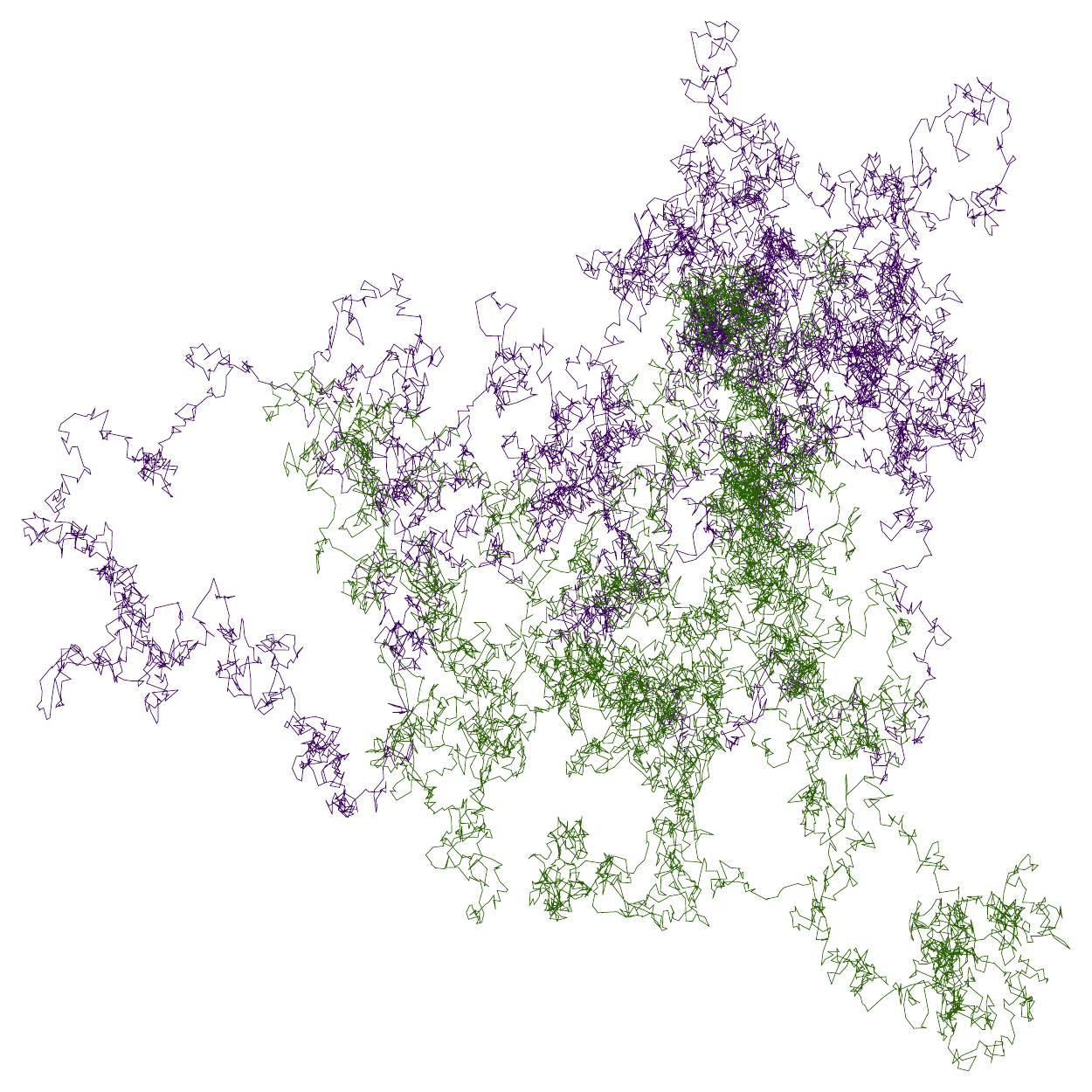}
        \caption{Two Brownian loops}
        \label{loop1}
    \end{subfigure}~
    \begin{subfigure}[b]{0.45\textwidth}
        \includegraphics[width=\textwidth]{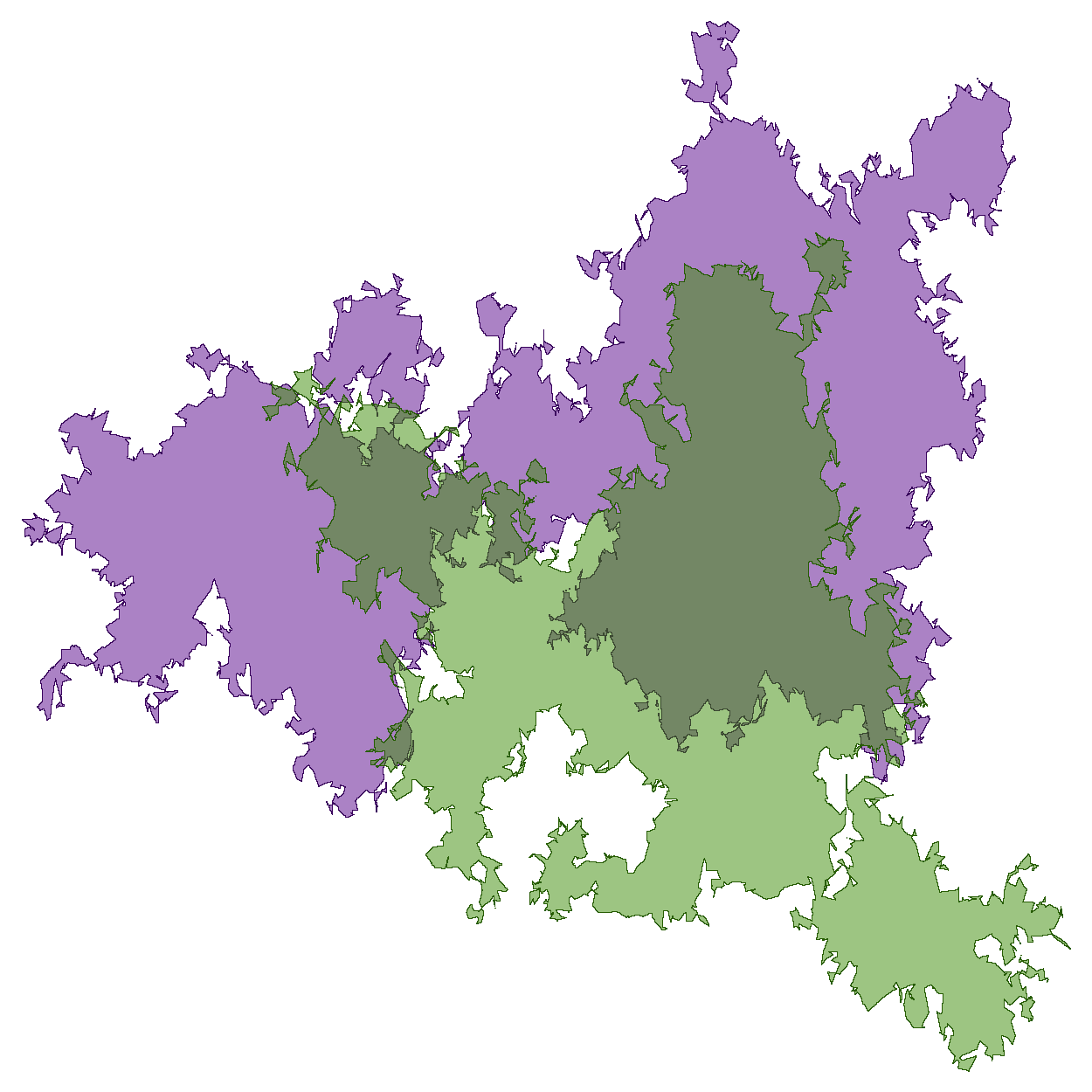}
        \caption{The outer boundaries of the same loops}
        \label{loop2}
    \end{subfigure}
    \caption{Two random Brownian loops  with identical parameters.  For $z$ within its outer boundary, each loop contributes $\pm 1$ to the layering number $N(z)$, where the sign is chosen uniformly randomly. If the purple loop is assigned $+1$ and the green loop $-1$, $N(z) = 0$ for $z$ in the white or dark green areas,  $+1$ in the purple, and $-1$ in the light green.  }
    \label{loop_figure}
\end{figure}

In this paper we will consider the case where $N(z)$ counts the layering number -- each loop that encircles the point $z$ contributes $\pm 1$ to $N(z)$, with the sign chosen uniformly  at random (see Fig.\ \ref{loop_figure}). 
Note that these layering vertex operators are sensitive only to the outer boundary of each Brownian loop (Fig.\ \ref{loop2}).

Because the BLS is a Poissonian ensemble of loops (that is, each loop is independent of the others), the probabilities of events in the BLS can be expressed in terms of $\mu^{\text{loop}}$. Specifically, we can obtain certain correlation functions in the BLS  with suitable cutoffs  simply by taking the exponential of $\lambda$ times the $\mu^{\text{loop}}$-weights of certain sets of loops.

In \cite{camia2016brownian} we demonstrated that,  after removing the cutoff, $e^{i \beta N(z)}$ is a scalar primary with conformal weights that are  periodic functions of $\beta$, $\Delta = \bar \Delta = \frac{\lambda}{10}(1 - \cos \beta)$, and that correlation functions of products of these operators vanish unless $\sum_i \beta_i = 0$ mod $2 \pi$. We computed the two- and three-point functions, but only up to multiplicative constants.  

 In this paper we  use the results of \cite{han2017brownian}
to  determine various expressions for the two- and three-point correlation functions. Together with a result of \cite{Gamsa_2006}, we use these results to compute analytically and in closed form the full four-point correlation function of the layering vertex operators in the plane, as a function of the positions of the four points, the intensity $\lambda$, and the four charges $\beta_i$.

The results of \cite{Gamsa_2006} were obtained by taking the limit $n \to 0$ of the critical $O(n)$ model, which is conformally invariant and known to describe self-avoiding loops in this limit.  As just mentioned, the layering vertex operators in the BLS are sensitive only to the outer boundary of each Brownian loop, and the outer boundary is by definition self-avoiding.  Furthermore, a result of \cite{werner2005conformally} guarantees that there is a unique (up to an overall multiplicative constant) conformally invariant measure on  self-avoiding loops (this is in fact the measure induced by $\mu^\text{loop}$ on the outer boundaries of Brownian loops).

With the four-point function in hand, we can expand it in conformal blocks. This reveals a new set of previously unknown conformal primary operators and their three-point function coefficients with the layering vertex operators. The physical interpretation and meaning of these operators is left to future work.

Using the results of \cite{han2017brownian} we  also compute the two-point function in the upper half-plane (subject to a certain boundary condition on the real axis), as a function of the positions of the two points, the intensity $\lambda$, and the two charges $\beta_i$.  The results of \cite{han2017brownian} are rigorous and based on SLE theory \cite{Beliaev13}.  They do not rely on the $n \to 0$ limit of the $O(n)$ model used in \cite{Gamsa_2006}.  Limits of this two-point function  help determine various constants in the four-point function.  In particular, we obtain the interesting result that the three-point function coefficient for three canonically normalized layering vertex operators is exactly 1, consistent with the conformal block expansion of the four-point function.  

We also determine the weights for Brownian loops to wind around one point and not another in the upper half-plane and full plane, and for several other configurations.

\section{Summary and results}
Our main results are the derivation of new correlation functions of exponentials of the layering operators in the BLS. In \cite{camia2016brownian},  three of us showed that the conformal dimensions of the operators $e^{i \beta N(z)}$ are 
\begin{align}
    \Delta = \bar \Delta = \frac{\lambda}{10}(1 - \cos \beta).
\end{align}
In this work we obtain the two-point function  of these operators in the upper half-plane with the boundary condition that any loop intersecting the real axis is erased, and the four-point function in the full plane.

The  two-point function in the upper half-plane $\mathbb{H}$ (Sec.\ \ref{sec2ptup}) is given by
\begin{align}
\begin{split}
\Braket{\tilde {\mathcal O}_{ \beta_1}(z_1) \tilde {\mathcal O}_{\beta_2} (z_2)}_{\mathbb{H}}
    = &|z_1-z_2|^{-2(\Delta_1+\Delta_2-\Delta_{12})} |z_1-\overline z_2|^{2(\Delta_1+\Delta_2-\Delta_{12})} |z_1-\overline z_1|^{-2\Delta_1}  |z_2-\overline z_2|^{-2\Delta_2} \\
    &\times \exp\left[ -(\Delta_1+\Delta_2- \Delta_{12})(1-\sigma) {}_3 F_{2}\left(1,1,\frac{4}{3};2,\frac{5}{3};  1-\sigma \right) \right],
\end{split}
\end{align}
 where $\tilde {\mathcal O}_{ \beta}(z) \propto e^{i \beta N(z)}$ are exponentials of layering operators normalized so that $\Braket{\tilde {\mathcal O}_\beta(z)}_{\mathbb{H}} = |z - \bar{z}|^{-2 \Delta}$, and $\sigma$, $\Delta_i,$ and $\Delta_{ij}$ are defined in \eqref{sigmadef} and \eqref{Ddef}.

The four-point function of these operators in the full plane ${\mathbb C}$  (Sec.\ \ref{sec4p}) is given by
\begin{align}
\begin{split}
    \Braket{\prod_{i=1}^4 {\mathcal O}_{ \beta_i}(z_i) }_{\mathbb C} =
    \exp\left[-\lambda A(x) \left( \sum_{i=1}^4 \Delta_i - \sum_{j=2}^4 \Delta_{1 j} \right) \right]
    \left|\frac{z_{13} z_{24}}{z_{34}z_{12}}\right|^{- \Delta_{12}}
   \left|\frac{z_{13} z_{24}}{z_{14}z_{23}}\right|^{- \Delta_{14}} \\ 
    \times \left |\frac{z_{12} z_{14}}{z_{24}}\right|^{- \Delta_1} 
   \left |\frac{z_{12} z_{23}}{z_{13}}\right|^{- \Delta_2} 
   \left|\frac{z_{23} z_{34}}{z_{24}}\right|^{- \Delta_3} 
   \left|\frac{z_{14} z_{34}}{z_{13}}\right|^{- \Delta_4},
\end{split}
\end{align}
with
\begin{align}
\begin{split}\label{intro4pt}
    A(x) = \frac{1}{4} \left[ x ~ {}_3 F_{2}\left(1,1,\frac{4}{3};2,\frac{5}{3}; x\right)
    +\overline x ~ {}_3 F_{2}\left(1,1,\frac{4}{3};2,\frac{5}{3}; 
    \overline x\right)  \right] \\
    \quad - \frac{2 \cdot 2^{\frac{1}{3}} \pi^2}{\sqrt{3}\G{\frac 1 6}^2
    \G{\frac 4 3}^2} | x (1-x) |^{\frac{2}{3}} \left|
    {}_2 F_{1}\left(\frac{2}{3},1:\frac{4}{3}; 
    x \right)\right|^2,
\end{split}
\end{align}
where the operators are normalized so that
$\Braket{{\mathcal O}_{\beta_1}(z_1) {\mathcal O}_{\beta_2}(z_2)}_{\mathbb C} = |z_1 - z_2|^{-2 \Delta_1}$ and $x$ is the cross-ratio \eqref{eq 10.8}.

All $n$-point functions in the full plane vanish unless a (periodic) charge conservation condition is satisfied:
\begin{align}
    \sum_{i=1}^n \beta_i = 2 \pi k, \,\,\,\, k\in \mathbb Z.
\end{align}

The conformal block expansion of the four-point function \eqref{intro4pt} (Sec.\ \ref{secconfbl}) reveals the spectrum of conformal primaries and associated three-point function coefficients. We find an apparently infinite new set of primary operators of integer spin in the BLS, with conformal dimensions
\begin{align}
\begin{split}
    \Delta^{(p,p')} &= \frac{\lambda}{10}(1-\cos(\beta_1+\beta_2)) + \frac{p}{3}\\
    \bar{\Delta}^{(p,p')} &= \frac{\lambda}{10}(1-\cos(\beta_1+\beta_2)) + \frac{p'}{3},
\end{split}
\end{align}
where $p$ and $p'$ are non-negative integers satisfying $p-p' = 0$ mod $3$.

Weights for Brownian loops to encircle various subsets of points in the plane or upper half-plane can be found in \eqref{eq 6.1}, \eqref{eq 8}, and \eqref{weights}.

\subsection{Motivation and previous work}

In \cite{Freivogel:2009rf}, Freivogel and Kleban  considered a toy model intended to capture the late-time physics of cosmic bubble nucleation in eternally inflating or de Sitter spacetime.  In a spacetime with one time and two space dimensions, these bubbles will be disks (with random fluctuations to their shape) that expand exponentially after their nucleation due to the expansion of the ambient spacetime.  This turns out to imply that on a late time slice the distribution of disks will be invariant under translations, scale transformations, rotations, and special conformal transformations \cite{Freivogel:2009it}. This ``disk soup'' has intensity $\lambda_\text{FK}$ that is equal to the bubble production rate per Hubble time per Hubble volume.  

It is widely believed that theories with Poincar\'e and scale invariance are fully conformally invariant.  However, the disk soup  model of \cite{Freivogel:2009rf} appears to be an exception.  Operators of the form $e^{i \beta N(z)}$, where $N(z)$ now counts the number of disks that overlap the point $z$ (the ``layering'' operator in the parlance of this paper) exhibit the behavior of  primary operators with dimension $\Delta = \bar \Delta = \frac{\pi}{2} \lambda_\text{FK} (1 - \cos \beta)$ -- explicit computation shows that their two- and three-point functions have the appropriate $z$-dependence. However, the four-point function, while crossing symmetric, is a non-analytic function of the $z_i$.  For this reason there is no conformal block expansion.  Presumably, this is because general conformal transformations do not map disks into disks, so the disk distribution is not invariant under local conformal transformations.  

A primary motivation for \cite{camia2016brownian} was to  obtain a full-fledged conformal field theory by replacing the disk distribution of \cite{Freivogel:2009rf} with the Brownian Loop measure \cite{werner2005conformally}. The analog of the disk model with the disk distribution replaced by the Brownian loop measure is precisely the BLS \cite{lawler2003brownian}. Since the BLS is known to be locally conformally invariant, a theory defined by it should be a full-fledged local conformal field theory.  While \cite{camia2016brownian} demonstrated that the exponentials of the (loop) layering operators are conformal primary operators, we were unable to compute the three-point function coefficients or  four-point correlation functions.  

In this work we take a major step beyond \cite{camia2016brownian} by obtaining explicit results for  the four-point function in the plane -- which indeed is an analytic function of the cross ratio and has a conformal block expansion -- as well as  the three-point function coefficients, and the two-point function on the upper half-plane.

\section{The two-point function in the upper half-plane}\label{sec2ptup}

In this section we will use the results of \cite{camia2016brownian,han2017brownian} to derive the general two-point function of layering vertex operators in the upper half-plane $\mathbb{H}$.  In this section and everything that follows, we will make extensive use of a main result from \cite{camia2016brownian}
\begin{align}\label{npoint}
    \Braket{\prod_{j=1}^n
    e^{i \beta_j N(z_j)}
    } = &\prod_{S \in \{z_1, \ldots\, z_n \}} \exp \left[
    - \lambda \,  \alpha(S|S^c) \left(1 - \cos\sum_{k \in I_S} \beta_k  \right) \right].
\end{align}
Here the product is over all  nonempty subsets $S \subset \{ z_1, \ldots, z_n \}$  and $I_S$ denotes the set of indices corresponding to the points of $\{ z_1, \ldots, z_n \}$ contained in $S$. The $\alpha(S|S^c)$ are the weights, according to the Brownian loop measure, of the sets of loops that encircle the points in $S$ but not those in $S^c$. The loops need to be contained in some domain $D$, which in this paper is either the upper half-plane or the full plane.
We will denote weights in the upper half-plane by $\alpha_\mathbb{H}$ and weights in the full plane simply by  $\alpha$, and correlation functions by $\braket{\ldots}_\mathbb{H}$ and $\braket{\ldots}_\mathbb{C}$, respectively.
In words, \eqref{npoint} states that a general $n$-point function of layering vertex operators in the BLS equals the exponential of terms consisting of the weights for loops that encircle various subsets of the points times the associated conformal weights.

The two-point function in any simply connected domain of the plane can be obtained from $\mathbb{H}$ by a conformal transformation, so computing the two-point function in $\mathbb{H}$  in principle gives the two-point function in a domain of any shape.  The boundary condition is that all loops must be confined entirely to the interior of $\mathbb{H}$ (that is, one could consider the BLS on the full plane  and remove all loops that intersect the lower half-plane).

Our strategy is to first find the weights of loops that encircle one or both points in $\mathbb{H}$.  Once we have these weights we can immediately write down the two-point function for general $\beta_i$ using \eqref{npoint}. 
We adopt a notation related to that of \cite{camia2016brownian}.
For two points $z_1 = x_1 + i y_1, z_2 = x_2+i y_2 \in \mathbb C$  let
\begin{align} \label{weightdef}
\begin{split}
    \alpha_{\mathbb{H}}(z_1|z_2) &= \mu^{\text{loop}}\{\gamma: \operatorname{diam}(\gamma) >\delta,
    \gamma \subseteq \mathbb{H}, z_1\in \overline \gamma, z_2 \not\in \overline \gamma\} \\ 
    \alpha_{\mathbb{H}}(z_1,z_2) &= \mu^{\text{loop}}\{\gamma: \operatorname{diam}(\gamma) >\delta,
    \gamma \subseteq \mathbb{H}, z_1, z_2\in \overline \gamma\}.
\end{split}
\end{align}
Here  $\delta>0$ is a short-distance regulator that we will later take to zero, $\gamma$ is a loop (left panel of Fig.\ \ref{loop_figure}),  $\overline \gamma$ is the interior of $\gamma$ (right panel of Fig.\ \ref{loop_figure}, shaded region), and $\operatorname{diam}(\gamma)$ is its diameter (the largest distance between any two points on the loop).

In general, the weights of loops that encircle only one point (such as \eqref{weightdef}, first line) diverge as $\delta \to 0$ due to contributions from arbitrarily small loops, infinitely many of which encircle any given point. Weights of loops that encircle two or more points (such as \eqref{weightdef}, second line) are finite as $\delta \to 0$ because only loops whose diameter is larger or equal to the distance between the two closest points encircle them.

For $|z_1-z_2|\geq \delta$ we have from \cite{han2017brownian}
\begin{align}
 \alpha_{ \mathbb{H}}(z_1,z_2) &= -\frac{\pi}{5 \sqrt{3}} -\frac 1{10} \eta {}_3 F_{2}\left(1,1,\frac{4}{3};2,\frac{5}{3}; \eta\right) -\frac 1{10} \log(\eta (\eta-1)) \label{eq 1}\\
 & \quad \quad + \frac{\Gamma(\frac 2 3)^2}{5 \Gamma(\frac 4 3)}(\eta(\eta-1))^{\frac{1}{3}}{}_2 F_{1}\left(1,\frac 2 3;\frac{4}{3}, \eta\right) \nonumber \\
 &=-\frac{1}{10}\left[\log\sigma+(1-\sigma) {}_3 F_{2}\left(1,1,\frac{4}{3};2,\frac{5}{3};  1-\sigma \right) \right],
  \label{eq 2}
\end{align}
where 
\begin{eqnarray}
\eta=-\frac{ (x_1-x_2)^2+(y_1-y_2)^2 }{ 4 y_1 y_2  } 
=\frac{ (z_1-z_2)(\overline z_1- \overline z_2) }{ (z_1-\overline z_1)(z_2- \overline z_2)  }
\end{eqnarray}
and
\begin{eqnarray} \label{sigmadef}
\sigma=\frac{ |z_1-z_2|^2 }{  |z_1-\overline z_2|^2  } =
\frac{ (x_1-x_2)^2+(y_1-y_2)^2 }{  (x_1-x_2)^2+(y_1+y_2)^2   } ;
\end{eqnarray}
notice that
$\eta-1
=\frac{ (z_1-\overline z_2)(\overline z_1- z_2) }{ (z_1-\overline z_1)(z_2- \overline z_2)  }$.

We can use \eqref{eq 2} and properties of the weight of the loops around $z_1$ in the Brownian loop measure to get
an expression for $ \alpha_{\mathbb{H}}(z_1|z_2)$.
Let 
\begin{align}
\begin{split}
 \alpha_{\delta,\mathbb{H}}(z_1)&=\mu^{\text{loop}}\{\gamma: \text{diam}(\gamma) >\delta,
\gamma \subseteq \mathbb{H}, z_1\in \overline \gamma\} \\
 \alpha_{\delta,{R}}(z_1)&=\mu^{\text{loop}}\{\gamma: R \geq \text{diam}(\gamma) >\delta,
 z_1\in \overline \gamma\}.
\end{split}
\end{align}
Then, by scale invariance of the Brownian loop measure, and Lemma A1 of \cite{camia2016brownian}, if $\delta \leq y_1$
\begin{eqnarray}
 \alpha_{\mathbb{H}}(z_1)=
 \alpha_{\delta,y_1}(z_1)+  \alpha_{y_1, \mathbb{H}}(z_1)
 =\frac 1 5 \log\frac {y_1}{\delta} + \overline \alpha
  =\frac 1 5 \log\frac {|z_1 -\overline z_1|  }{2\delta} + \overline \alpha,
\end{eqnarray}
where $\overline \alpha = \alpha_{1, \mathbb{H}}(i)$, the weight of the loops around the
point $z=i$ with diameter greater than or equal to $1$ and contained in $\mathbb{H}$, 
is a constant of the model.

On the other hand, if $|z_1-z_2| \geq \delta$, then
\begin{eqnarray}\label{eq 5.1}
 \alpha_{\delta, \mathbb{H}}(z_1)=
 \alpha_{\mathbb{H}}(z_1|z_2)+  \alpha_{ \mathbb{H}}(z_1, z_2);
\end{eqnarray}
hence, for $\delta\leq \min(|z_1-z_2|, |z_1-\overline z_1| / 2 )$,
 \begin{eqnarray} \label{eq 6}
  \alpha_{\mathbb{H}}(z_1|z_2)=
  - \alpha_{ \mathbb{H}}(z_1, z_2)
+\frac 1 5 \log|z_1-\overline z_1|-\frac 1 5 \log(2\delta) + \overline \alpha.
\end{eqnarray}

Using \eqref{npoint} 
and denoting  the conformal dimensions by
\begin{align}
\begin{split}\label{Ddef}
    \Delta_j &=\frac{\lambda}{10} (1-\cos\beta_j) \\
    \text{and}\quad \Delta_{i j} &= \frac{\lambda}{10}(1-\cos(\beta_i+\beta_j)),
    \end{split}
\end{align}
we have
\begin{align}
\begin{split}\label{eq 4}
  &\Braket{e^{i \beta_1 N(z_1)} e^{i \beta_2 N(z_2)}}_\mathbb{H} \\
  =& \exp \left[ -\lambda\left( (1-\cos\beta_1)  \alpha_{ \mathbb{H}}(z_1|z_2)
  + (1-\cos\beta_2)  \alpha_{\mathbb{H}}(z_2|z_1)  
   + (1-\cos(\beta_1+\beta_2))  \alpha_{ \mathbb{H}}(z_1, z_2)  \right)
  \right] \\
    =& \left(2\delta e^{-5 \overline \alpha}\right)^{2 (\Delta_1+\Delta_2)} |z_1-z_2|^{-2(\Delta_1+\Delta_2-\Delta_{12})}
     |z_1-\overline z_2|^{2(\Delta_1+\Delta_2-\Delta_{12})} |z_1-\overline z_1|^{-2\Delta_1}  |z_2-\overline z_2|^{-2\Delta_2} \\
  &   \quad 
  \times \exp\left[ -(\Delta_1+\Delta_2- \Delta_{12})(1-\sigma) {}_3 F_{2}\left(1,1,\frac{4}{3};2,\frac{5}{3};  1-\sigma \right) \right]
\end{split}
\end{align}
Notice that the equation suggests a specific ultraviolet cutoff in this ``renormalization scheme'': in addition to the normalizing factor $\delta^{2 \Delta_j}$
used in  \cite{camia2016brownian}, it is natural to include the constants
$(2 e^{-5 \overline \alpha})^{2 \Delta_j}$. Defining $
\tilde{\mathcal O}_\beta(z) \equiv \left({2 \delta e^{-5 \bar\alpha}  }\right)^{-2\Delta}e^{i \beta N(z)},
$ the two-point function  becomes
\begin{align}
\begin{split}\label{eq 3}
    \Braket{\tilde {\mathcal O}_{ \beta_1}(z_1) \tilde {\mathcal O}_{\beta_2} (z_2)}_{\mathbb{H}}
    \equiv& \lim_{\delta \rightarrow 0} \frac{ \Braket{e^{i \beta_1 N(z_1)} e^{i \beta_2 N(z_2)}}_{\mathbb{H}}} {\left(2 \delta e^{-5 \overline \alpha}\right)^{2 (\Delta_1+\Delta_2)}} \\
    =& |z_1-z_2|^{-2(\Delta_1+\Delta_2-\Delta_{12})} |z_1-\overline z_2|^{2(\Delta_1+\Delta_2-\Delta_{12})} |z_1-\overline z_1|^{-2\Delta_1}  |z_2-\overline z_2|^{-2\Delta_2} \\
    & \exp\left[ -(\Delta_1+\Delta_2- \Delta_{12})(1-\sigma) \, {}_3 F_{2}\left(1,1,\frac{4}{3};2,\frac{5}{3};  1-\sigma \right) \right].
\end{split}
\end{align}
Note that the one-point function can be immediately obtained from \eqref{eq 3} by setting one of the $\beta_i = 0$:
\begin{align}
    \Braket{\tilde {\mathcal O}_{ \beta_1}(z_1) }_{\mathbb{H}}
    =&  |z_1-\overline z_1|^{-2\Delta_1}.
\end{align}
Using the equivalence  \eqref{eq 1}, \eqref{eq 3} can also be written as
\begin{align}
\begin{split}
    &|z_1-z_2|^{-2(\Delta_1 + \Delta_2 - \Delta_{12})} |z_1-\overline z_2|^{-2(\Delta_1 + \Delta_2 - \Delta_{12})} |z_1-\overline z_1|^{-2(\Delta_2 - \Delta_{12})} |z_2-\overline z_2|^{-2(\Delta_1 - \Delta_{12})}\\
    &\exp \Bigg[ 2(\Delta_1+\Delta_2-\Delta_{12}) \\
    & \times \left( -\frac{\pi}{\sqrt{3}} -\frac{1}{2} \eta {}_3 F_{2}\left(1,1,\frac{4}{3};2,\frac{5}{3}; \eta \right) 
     + \frac{\Gamma(\frac 2 3)^2}{5 \Gamma(\frac 4 3)} (\eta(\eta-1))^{\frac{1}{3}} {}_2 F_{1}\left(1,\frac 2 3;\frac{4}{3}, \eta \right) \right) \Bigg].
\end{split}
\end{align}
The results obtained above are the two-point functions of the field obtained after renormalization and the limit $\delta \to 0$, whose existence in Sobolev spaces $\mathcal{H}^{-\alpha}(\mathbb{H}), \alpha>3/2$, is shown in \cite{camia2019brownian} for $\Delta < 1/2$.

\section{The two- and three-point functions in the full plane}\label{sec23full}

In this section we compute the two- and three-point functions of layering vertex operators in the full plane.  
A major difference from the half-plane is that in the full plane
all correlation functions go to zero because of the contribution from large loops, unless the charge conservation condition
\begin{align}\label{cc}
    \sum_{i=1}^n \beta_i=2 \pi  k, \,\,\,\, k\in \mathbb Z
\end{align}
is satisfied \cite{Freivogel:2009rf, camia2016brownian}. This is reminiscent of momentum or charge conservation for the vertex operators of the free boson, where the condition arises from integration over the zero mode.  

Note that this condition would be clearly necessary were we to define these correlation functions on the sphere rather than the plane, because on a sphere a loop that covers a subset of points can equally well be interpreted as a loop that covers the complement of that set (on a compact space there is no notion of the ``inside'' versus the ``outside'' of the loop).
Since the plane and the sphere are conformally equivalent, \eqref{cc} is necessary for consistency (\emph{cf.} Sec.\ \ref{sec4p}).

The $z_i$ dependence of the two- and three-point functions in the full plane follow from conformal invariance and the fact that the layering vertex operators are conformal primaries \cite{camia2016brownian}. However, this argument does not fix the constant prefactors, which were not computed in \cite{camia2016brownian}.
As we show in Appendix \ref{appendix}, by taking the limit that the points are far from the boundary, we can use our results from the upper half-plane to determine the multiplicative prefactors left undetermined in \cite{camia2016brownian}.
We find that the most convenient normalization in the plane is
\begin{equation}\label{planenorm}
\mathcal O_\beta(z) \equiv \left({2 \delta e^{-\frac{\pi}{\sqrt{3}}-5 \bar\alpha}  }\right)^{-2\Delta}e^{i \beta N(z)},
\end{equation}
 where again $\bar \alpha\equiv \alpha_{1,\mathbb{H}}(i)$ is a constant equal to the weight of loops in the upper half-plane with a diameter larger than 1 and that encircle $z=i$.  Comparing to \eqref{eq 3}, the difference between the canonically normalized layering vertex operator in the upper half-plane $\tilde {\mathcal O}$ and the full plane normalization in \eqref{planenorm} is simply the factor $e^{\frac{2 \pi}{\sqrt3} \Delta}$.
 
 With this definition the two- and three-point functions in the limit $\delta \to 0$ are
\begin{equation}
    \Braket{\mathcal O_{\beta_1}(z_1) \mathcal O_{\beta_2}(z_2)}_{\mathbb C} = |z_1 - z_2|^{-4 \Delta_1}
\end{equation}
and
\begin{equation} \label{24}
\Braket{\mathcal O_{\beta_1}(z_1) \mathcal O_{\beta_2}(z_2)O_{\beta_3}(z_3)}_{\mathbb C}=|z_1-z_2|^{-2(\Delta_1+\Delta_2-\Delta_{3})}
     |z_1- z_3|^{-2(\Delta_1+\Delta_3-\Delta_{2})}
     |z_2- z_3|^{-2(\Delta_2+\Delta_3-\Delta_{1})};
\end{equation}
  the calculation of \eqref{24} is possible
  since, when the three-point function of the fields is expressed in terms of  $\alpha(z_1,z_2)$ and $\alpha(z_1,z_2, z_3)$, the net
contribution of the last of these two terms vanishes by charge conservation
(Appendix \ref{appendix}).
  Remarkably, the three-point function coefficients are precisely 1 for all values of the $\beta_i$ satisfying \eqref{cc}.

In the notation of \cite{camia2016brownian}, these results are equivalent to
\begin{equation}
    C_2=\left(2 e^{-\frac{\pi}{\sqrt{3}}-5 \bar\alpha}\right)^{2(\Delta_1+\Delta_{2})},
    \quad C_3 =\left(2 e^{-\frac{\pi}{\sqrt{3}}-5 \bar\alpha}\right)^{2(\Delta_1+\Delta_{2}+\Delta_{3})}.
\end{equation}

\section{Nacu-Werner thinness function}

Starting from \eqref{eq 6} and \eqref{eq 2} and a small ultraviolet cutoff constant $\delta$, one can get
an explicit expression of the Nacu-Werner thinness function $\alpha(z_1|z_2)$ (the weight of loops that encircles one point but not another) in terms of $\overline \alpha$.
The fact that this function is finite appeared in \cite{Nacu_2011}.
In fact, for two points $z_1, z_2 \in \mathbb C$,
$\delta\leq |z_1-z_2|$, and 
$z_1(t), z_2(t)$ as in the previous section, we have
\begin{align}\label{eq 6.1}
\begin{split}
    &\alpha(z_1|z_2) = \alpha(z_2|z_1) = \lim_{t \to \infty} \alpha_{\mathbb{H}}(z_1(t)|z_2(t)) \\
    =& \lim_{t \to \infty} \left[ - \alpha_{ \mathbb{H}}(z_1(t), z_2(t)) +\frac 1 5 \log|z_1(t)-\overline z_1(t)|-\frac 1 5 \log(2\delta) + \overline \alpha \right] \\
    =& \frac{1}{5} \left[ \log |z_1-z_2| + \log \frac{|z_1-\overline z_1|^2}{|z_1- z_1|^2} + (1-\sigma) {}_3 F_{2}\left(1,1,\frac{4}{3};2,\frac{5}{3};  1-\sigma \right)\right] - \frac{1}{5} \log (2 \delta) + \overline \alpha  \\
    =& \frac{1}{5} \log |z_1-z_2| + Q,
\end{split}
\end{align}
where we defined the constant
\begin{align}\label{Q}
    Q = \frac{\pi}{5 \sqrt{3}} - \frac{1}{5} \log(2 \delta) +  \bar \alpha
\end{align}
and again $\bar \alpha \equiv \alpha_{1,\mathbb{H}}(i)$ is the weight of loops in the upper half-plane with a diameter larger than 1 and that encircle $z=i$.  This is the Nacu-Werner function with ultraviolet cutoff $\delta\leq |z_1-z_2|$, which turns out to be the fundamental solution of the Laplacian
in $\mathbb C$, except for one multiplicative and one additive constant. 
Note that the analogous calculation of $\alpha(z_1|z_2, z_3)$
from \eqref{4.10}  is not possible as $\alpha(z_1, z_2, z_3)$ is not known.

On the other hand, by combining \eqref{eq 6} with \eqref{eq 1}, 
we get an explicit expression for the linear term of the
$O(n)$ expansion in \cite{Gamsa_2006}, which confirms
their calculation, except for a factor of $6 \pi/5$,
to correct for different scaling, and the precise form of the constant. In fact
\begin{align} \label{eq 8}
\begin{split}
 & \alpha_{\mathbb{H}}(z_1|z_2)
  +\alpha_{\mathbb{H}}(z_2|z_1) \\
  =& - 2\alpha_{ \mathbb{H}}(z_1(t), z_2(t))
+\frac 1 5 \log(|z_1(t)-\overline z_1(t)||z_2(t)-\overline z_2(t)|) -\frac 2 5 \log(2\delta) + 2\overline \alpha\\
 =& \frac{2\pi}{5 \sqrt{3}} +\frac 1{5} \eta {}_3 F_{2}\left(1,1,\frac{4}{3};2,\frac{5}{3}; \eta\right)
 +\frac 1{5} \log(\eta (\eta-1)) - \frac{2 \Gamma\left(\frac 2 3 \right)^2}{5 \Gamma(\frac 4 3)}(\eta(\eta-1))^{\frac{1}{3}} {}_2 F_{1}\left(1,\frac 2 3;\frac{4}{3}; \eta\right) \\
 & \quad +\frac 1 5 \log(|z_1(t)-\overline z_1(t)||z_2(t)-\overline z_2(t)|)
-\frac 2 5 \log(2\delta) + 2\overline \alpha\\
 =& -\frac 1 5 \bigg[-\eta {}_3 F_{2}\left(1,1,\frac{4}{3};2,\frac{5}{3}; \eta\right)+ \frac{2 \Gamma\left(\frac 2 3 \right)^2}{ \Gamma\left(\frac 4 3 \right)}(\eta(\eta-1))^{\frac{1}{3}} {}_2 F_{1}\left(1,\frac 2 3;\frac{4}{3}; \eta\right) \\
 & \quad -\log(\eta (z_1-\overline z_2) (z_2 - \overline z_1))
 \bigg] + 2 Q,
\end{split}
\end{align}
equals the expression in \cite{Gamsa_2006}, bottom of Page 18, multiplied by $6 \pi/5$.

This confirms the validity of the $O(n)$ expansion in \cite{Gamsa_2006}
for the two point function. In the next section we use
the same expansion for the four-point function, this time
without  an independent verification.

\section{The general four-point function in the plane}\label{sec4p}

In this section we will compute the general four-point function for the layering operators in the whole plane.  As for the two-point function in the upper half-plane, we will first derive the weights for loops covering various subsets of the points, and then with these in hand we can immediately write down the correlation function using \eqref{npoint}. One difference is that to derive the weights we  rely on a result of \cite{Gamsa_2006}.

Consider four points $z_1,z_2,z_4,z_4$ and assume in what follows that the letters $i,j,k,\ell \in \{ 1,2,3,4 \}$ are always different. Using \eqref{npoint}, the four-point function is
\begin{align}
\begin{split}\label{4pt}
\Braket{ \prod_{i=1}^4 e^{i \beta_i N(z_i)} }_\mathbb{C}
= \exp \Bigg[ -\lambda \Bigg(
    &\sum_{i=1}^4(1-\cos\beta_i) \alpha(z_i|z_j, z_k, z_{\ell}) \\
    +& \sum_{\substack{i,j=1\\i<j}}^4(1-\cos(\beta_i+\beta_j)) \alpha(z_i, z_j| z_k, z_{\ell}) \\
    +& \sum_{i=1}^4 (1-\cos(\beta_j+\beta_k+\beta_\ell)) \alpha(z_j,z_k,z_{\ell}|z_i)\\
    +& (1-\cos(\beta_1+\beta_2+\beta_3+\beta_4)) \alpha(z_1,z_2,z_3,z_4)
    \Bigg) \Bigg].
\end{split}
\end{align}
The weights of loops encircling points in the full plane are defined analogously to \eqref{weightdef}.

Given we satisfy the charge conservation condition \eqref{cc}, the four-point function is independent of $\alpha(z_1,z_2,z_3,z_4)$ and we can rearrange the terms as follows:
\begin{align}
\begin{split}\label{4ptac}
\Braket{ \prod_{i=1}^4 e^{i \beta_i N(z_i)} }_\mathbb{C} 
= \exp \Bigg[ - \lambda \Bigg(
    &\sum_{i=1}^4(1-\cos\beta_i) \ac(z_i|z_j, z_k, z_{\ell}) \\
    + &\sum_{j=2}^4(1-\cos(\beta_1+\beta_j)) \ac(z_1, z_j| z_k, z_{\ell})
    \Bigg) \Bigg],
\end{split}
\end{align}
where we introduced the weights
\begin{align}
    \ac(S | S^c) \equiv \alpha(S | S^c) + \alpha( S^c | S)
\end{align}
for subsets of points  $ S \subset \{z_1,z_2,z_3,z_4\}$, with $S^c$ the complement of $S$.  For instance, $\ac(z_1 |z_2,z_3,z_4) = \alpha(z_1 |z_2,z_3,z_4) + \alpha(z_2,z_3,z_4 | z_1)$.

As previously mentioned, if we consider the BLS on a sphere rather than the plane, charge conservation is necessary for consistency because there is no distinction between the inside and outside of a loop on a sphere.  Another implication of this fact is that both ``sides'' of the loop must contribute equally to the correlation functions.  Under stereographic projection to the plane, the “outside” of the loop is the side that contains the point that projects to infinity of the plane, but it remains the case that both the inside and the outside must contribute.  This explains why only the paired weights $\ac$ appear in \eqref{4ptac}.

There are a total of seven pairs $\ac$ that contribute. Six of these can be determined from the results we have already obtained for the two-point functions (we can also obtain relations from the three-point functions, but they are not independent). For the seventh relation we will use a result of \cite{Gamsa_2006}.

Choosing $\beta_1=\beta_2=\pi, \beta_3=\beta_4=0$ in \eqref{4ptac} reproduces the two-point function \eqref{2pt}. Comparing these,  we  obtain the relation
\begin{align}\label{set1}
\begin{split}
    2 \alpha_\delta(z_1 | z_2) &= \ac(z_1|z_2, z_3, z_4) + \ac(z_2|z_1,z_3,z_4) + \ac(z_1, z_3|z_2,z_4) + \ac(z_1,z_4|z_2,z_3)\\
    &= \frac{2}{5} \log |z_1-z_2| + 2 Q,
\end{split}
\end{align}
where we used \eqref{eq 6.1} in the last line.
Five other independent equations can be obtained by choosing other pairs of the $\beta_i$ equal to $\pi$ and $0$. 

The system of six equations we obtain from  \eqref{set1} and its permutations has rank six.  An additional independent relation is necessary to solve for the seven $\ac$, and is provided by \cite{Gamsa_2006}. 
Defining $z_{jk}=(z_j-z_k)$, the cross-ratio
\begin{align}\label{eq 10.8}
    x = \frac{z_{12} z_{34}}{z_{13} z_{24}}, \quad 1- x = \frac{z_{14} z_{23}}{z_{13} z_{24}},
\end{align}
and the function
\begin{eqnarray}
A(x)&=& \frac{1}{4} \left[
x ~ {}_3 F_{2}\left(1,1,\frac{4}{3};2,\frac{5}{3}; x\right)
+\overline x ~ {}_3 F_{2}\left(1,1,\frac{4}{3};2,\frac{5}{3}; 
\overline x\right)  \right] \label{eq 10.9}\nonumber\\
&& \quad - \frac{2 \cdot 2^{\frac{1}{3}} \pi^2}{\sqrt{3}\G{\frac 1 6}^2
\G{\frac 4 3}^2} | x (1-x) |^{\frac{2}{3}} \left|
{}_2 F_{1}\left(\frac{2}{3},1;\frac{4}{3}; 
x \right)\right|^2,  \label{eq 10.10}
 \end{eqnarray}
 equations (21) and (22) of \cite{Gamsa_2006} imply that
  \begin{align}  \label{eq 10.11}
\sum_{i=1}^4 \alpha_{\mathbb S}(z_i|z_{j},  z_{k}, z_{\ell}) &=
P \left( \log |x z_{2 3} z_{1 4}|+
2 A(x) \right)+ 4 (Q + R).
 \end{align}
Cardy and Gamsa derived this result by solving a linear differential equation that does not fix the overall normalization or the additive constant, so we have included an overall coefficient $P$, and retained an additive constant that we denote $4 (Q + R)$ for future convenience (we will see shortly that $R=0$, and $Q$ is defined by \eqref{Q}).
 
 To determine $P$ we can examine the scaling behavior of the four-point function \eqref{4ptac} where we set $\beta_i=\pi$ and therefore $\Delta_i = \lambda/5$, $\Delta_{ij}=0$.  In general, if $O_{\Delta, \Delta}$ is a primary of dimension $(\Delta, \Delta)$, rescaling $z_i \to \rho z_i$ takes $\ln \Braket{O_{\Delta, \Delta}^4} \to -8 \Delta \ln \rho + \ln \Braket{O_{\Delta, \Delta}^4}$.  Using \eqref{4ptac}, this fixes $P=2/5$.

We now have seven independent equations for the seven $\ac$.  
The solutions are
\begin{subequations}\label{weights}
\begin{eqnarray}
	\alpha_{\mathbb S}(z_1|z_2,z_3,z_4) &=&
	\frac{1}{5}\left( \log\left|\frac{z_{12} z_{14}}{z_{24}}\right|
	+ A(x) \right) + Q + R \label{eq 10.12}\\
	 \alpha_{\mathbb S}(z_2|z_1,z_3,z_4) &=&
	\frac{1}{5}\left(\log\left|\frac{z_{12} z_{23}}{z_{13}}\right|
	+ A(x)\right) + Q + R \label{eq 10.13}\\
	 \alpha_{\mathbb S}(z_3|z_1,z_2,z_4) &=&
	\frac{1}{5}\left(\log\left|\frac{z_{23} z_{34}}{z_{24}}\right|
	+ A(x)\right) + Q + R \label{eq 10.14}\\
	 \alpha_{\mathbb S}(z_4|z_1,z_2,z_3) &=&
	\frac{1}{5}\left(\log\left|\frac{z_{14} z_{34}}{z_{13}}\right|
	+ A(x)\right) + Q + R \label{eq 10.15}\\
	 \alpha_{\mathbb S}(z_1,z_2|z_3,z_4) &=&
	-\frac{1}{5} \left( \log|x| + A(x) \right) - R \label{eq 10.16}\\
	 \alpha_{\mathbb S}(z_1, z_3|z_2,z_4) &=&
	-\frac{1}{5} A(x)  - R \label{eq 10.17}\\
	 \alpha_{\mathbb S}(z_1, z_4|z_2,z_3) &=&
	 -\frac{1}{5}\left( \log|1-x| + A(x) \right) - R \label{eq 10.18}
\end{eqnarray}
\end{subequations}

We can now show that $R=0$. 
Consider the four points $z_i$ arranged in a rectangle in cyclical order. If we let a pair of points approach the other pair by taking $z_1 \to z_4$ and $z_2 \to z_3$ it is clear that $\alpha(z_1,z_3|z_2,z_4) \to 0$,
since the probability of a loop passing between the pairs of points goes to zero.
In the same limit we have that $x \to 1,~ A(x) \to 0$.
Comparing this with \eqref{eq 10.17} shows that $R=0$.
The weights \eqref{eq 10.16}-\eqref{eq 10.18} coincide
with those given in (29)-(31) of \cite{Gamsa_2006}, after multiplication
by an overall factor ${6 \pi}/{5}$.

This allows us to write the fully general, normalized four-point function as
\begin{align}
\begin{split}\label{gen4pt}
    \Braket{\prod_{i=1}^4 {\mathcal O}_{\beta_i}(z_i)}_\mathbb{C} 
    &= \lim_{\delta \to 0} \left(2\delta e^{-5 \overline \alpha -\frac{\pi}{\sqrt 3}}\right)^{-2 \sum_{j=1}^4\Delta_j }
    \Braket{\prod_{i=1}^4 e^{i \beta_i N(z_i)} }_\mathbb{C}\\
    &= \exp\left[- 2 A(x) \left( \sum_{i=1}^4 \Delta_i - \sum_{j=2}^4 \Delta_{1 j} \right) \right]
    \left|\frac{z_{13} z_{24}}{z_{12}z_{34}}\right|^{-  2\Delta_{12}}
   \left|\frac{z_{13} z_{24}}{z_{14}z_{23}}\right|^{-  2\Delta_{14}} \\
   & \quad \times \left |\frac{z_{12} z_{14}}{z_{24}}\right|^{-2 \Delta_1}
   \left |\frac{z_{12} z_{23}}{z_{13}}\right|^{- 2\Delta_2} 
   \left|\frac{z_{23} z_{34}}{z_{24}}\right|^{- 2\Delta_3} 
   \left|\frac{z_{14} z_{34}}{z_{13}}\right|^{-2 \Delta_4},
\end{split}
\end{align}
where $A(x)$ is defined by \eqref{eq 10.10} and the $\Delta_i, \Delta_{ij}$ by \eqref{Ddef}.

With some algebra and using the identity 
\begin{align}\label{mysteriousident}
     A(x) - A(1/x) + \ln |x| = 0
\end{align}
one can check that the four-point function is invariant under exchange of any pair of indices, establishing crossing invariance.

\subsection{Free-field limit}

There is a limit in which the correlators in the full plane become those of free field vertex operators (the same limit was considered and the same result   obtained in \cite{Freivogel:2009rf}, for the disk model studied there). 
Consider taking $\beta_i \to 0$ and $\lambda \to \infty$ with the product $\lambda \beta_i^2 $  fixed. We define the field $\psi$ by $\beta N(z) = \sqrt{2} \gamma \psi(z)$ with
\begin{align}
    \gamma = \sqrt{\frac{\lambda}{20}} \beta
\end{align}
such that the conformal dimension of the operator $e^{i \beta N(z)} = e^{i \sqrt{2} \gamma \psi(z)}$ becomes
\begin{align}
    \Delta = \frac{\lambda}{10}(1 - \cos \beta) \to \frac{\lambda}{20} \beta^2 = \gamma^2.
\end{align}
This is the correct dimension for a canonically normalized free-field vertex operator $e^{i \sqrt{2} \gamma \psi}$.
Now consider \eqref{gen4pt} and note that
\begin{align}\label{cvan}
    \sum_{i=1}^4 \Delta_i - \sum_{j=2}^4 \Delta_{1 j} \to \sum_{i=1}^4 \gamma_i^2 - \sum_{j=2}^4 (\gamma_1 + \gamma_j)^2 = 0,
\end{align}
where we used the fact that $\sum_{i=1}^4 \gamma_i = 0$. Therefore the factor in \eqref{gen4pt} involving hypergeometric functions goes to 1 in this limit, and the remainder reduces immediately to the correct form for the four-point function of free-field vertex operators:
\begin{align}
    \Braket{\prod_{i=1}^4 {\mathcal O}_{\beta_i}(z_i)}_\mathbb{C} \to  \prod_{\substack{i,j=1\\i<j}}^4 |z_{i j}|^{4 \gamma_i \gamma_j} = \Braket{\prod_{j=1}^4 e^{i \sqrt{2} \gamma_j \psi(z_j)}}.
\end{align}
This same limit should reduce the $n$-point function in the plane for all $n$ to the free-field case. 

Interestingly, the correlators in the upper half-plane do \emph{not} reduce to those of free fields in the same limit.  To see this, note that the coefficient of the hypergeometric function in \eqref{eq 3}, $\Delta_1 + \Delta_2 - \Delta_{12}$, does not vanish in the limit described above.  (This is in contrast to the coefficient of the hypergeometric functions in \eqref{gen4pt}, see \eqref{cvan}.)  Since the two-point function of free-field vertex operators in the upper half-plane is simply a product of powers of distances between $z_1$ and $z_2$ and  their images in the lower half-plane $z_1^*$ and $z_2^*$,  \eqref{eq 3} does not reduce to the free-field result.  Apparently, the boundary condition on the real axis induces interactions between the bulk operators.  

\section{Expansion in conformal blocks}\label{secconfbl}

The four-point function of a conformal field theory contains information about the three-point function coefficients, as well as the spectrum of primary operators.
To obtain this data, one  makes use of the operator algebra by performing a conformal block expansion.

By a global conformal transformation, one can always map three of the points appearing in the four-point function to fixed values. The remaining dependence is only on the cross-ratio \eqref{eq 10.8}
\begin{align}
    x = \frac{z_{1 2} z_{3 4}}{z_{1 3} z_{2 4}} ,
\end{align}
and its conjugate $\bar{x}$. Each cross-ratio is invariant under global conformal transformations.
Following the notation of \cite{DiFrancesco:639405} Sec.\ $6.6.4$, we set $z_1 = \infty, ~z_2 = 1, ~z_3 = x$ and $z_4 = 0$,
and define
\begin{align}\label{defG1}
    G^{2 1}_{3 4}(x)  &= \lim_{z_1 \rightarrow \infty} z_1^{2 \Delta_1} \bar{z}_1^{2 \overline{\Delta}_1} 
    \Braket{ {\mathcal O}_{\beta_1}(z_1) {\mathcal O}_{\beta_2}(1) {\mathcal O}_{\beta_3}(x) {\mathcal O}_{\beta_4}(0)}_\mathbb{C}
\end{align}
 where $\overline{\Delta}_1={\Delta}_1$ in our case 
(note that later on we will consider operators with spin,
$\Delta^{(p,p')} \neq \bar \Delta^{(p,p')})$.

We can now proceed to expand the four-point function in Virasoro conformal blocks
\begin{align}\label{confblexp}
    G^{21}_{34}(x) = \sum_{\mathcal{P}} C_{34}^\mathcal{P} C_{12}^\mathcal{P} {\mathcal F}^{21}_{34}(\mathcal{P}|x)  \bar{\mathcal F}^{21}_{34}(\mathcal{P}|\bar x).
\end{align}
The sum over $\mathcal{P}$ runs over all primary operators in the theory, and the $C_{ij}^\mathcal{P}$ are the three-point function coefficients of the operators labeled by $i, j$ with $\mathcal{P}$. Each $\mathcal{P}$ with a non-zero $C$ contributes a term consisting of a holomorphic function times an anti-holomorphic function of the cross-ratio.  These functions -- the Virasoro conformal blocks -- depend only on $x$, the central charge $c$, and the conformal dimensions $\Delta_i, \Delta_\mathcal{P}$ of the five operators.

The conformal blocks are given perturbatively by a power series
\begin{align}\label{cblockexp}
    {\mathcal F}^{21}_{34}(\mathcal{P}|x) = x^{\Delta_\mathcal{P}-\Delta_3-\Delta_4}\sum_{K=0}^{\infty}{\mathcal F}_K x^K,
\end{align}
where coefficients $F_K$ are determined by the Virasoro algebra. 
The first 3 terms are given by
\begin{align}
    {\mathcal F}_0 &= 1 \\
    {\mathcal F}_1 &= \frac{(\Delta_\mathcal{P} +\Delta_2 - \Delta_1)(\Delta_\mathcal{P} + \Delta_3 - \Delta_4)}{2 \Delta_\mathcal{P}}\\
    {\mathcal F}_2 &= \frac{A + B}{C},
\end{align}
with
\begin{align}
\begin{split}
    A &= (\Delta_\mathcal{P} + \Delta_2 - \Delta_1)(\Delta_\mathcal{P} + \Delta_2 - \Delta_1 + 1) \\
    &\times [(\Delta_\mathcal{P} + \Delta_3 - \Delta_4)(\Delta_\mathcal{P} + \Delta_3 - \Delta_4 + 1)(4 \Delta_\mathcal{P} + c/2) - 6 \Delta_\mathcal{P} (\Delta_\mathcal{P} + 2 \Delta_3 - \Delta_4)]
\end{split} \\
    B &= 4 \Delta_\mathcal{P} (2 \Delta_\mathcal{P} + 1) (4 \Delta_\mathcal{P} + c/2) - 36 \Delta_\mathcal{P}^2 \\
\begin{split}
    C &= (\Delta_\mathcal{P} + 2 \Delta_2 - \Delta_1)\\
    &\times [4 \Delta_\mathcal{P} (2 \Delta_\mathcal{P} + 1)(\Delta_\mathcal{P} + 2 \Delta_3 - \Delta_4) - 6 \Delta_\mathcal{P} (\Delta_\mathcal{P} + \Delta_3 - \Delta_4)(\Delta_\mathcal{P} + \Delta_3 - \Delta_4 + 1)]
\end{split}
\end{align}
(see \cite{headrick, DiFrancesco:639405}).

We now take the limit  in the four-point function \eqref{gen4pt} to obtain
$G$:
\begin{align}\label{G1}
	G^{21}_{34}(x) &= |x|^{-2(\Delta_{12} - \Delta_3 - \Delta_4)} |1-x|^{-2(\Delta_{14} - \Delta_2 - \Delta_3)} \exp \left[ {2 \tilde{\Delta}} A(x) \right]
\end{align}
where $A(x)$ is given by \eqref{eq 10.10} and
\begin{align}
\begin{split}
	\tilde{\Delta}
	&= \Delta_{12} + \Delta_{13} + \Delta_{14} - \Delta_1 - \Delta_2 - \Delta_3 - \Delta_4, \\
	&= \frac{\lambda}{10}  \left[ -1 +\cos \beta_1 +\cos \beta_2 +\cos \beta_3 +\cos \beta_4 \right. \\
	&\quad\quad\quad\quad \left. - \cos(\beta_1 + \beta_2) - \cos(\beta_1 + \beta_3) - \cos(\beta_1 + \beta_4) \right].
\end{split}
\end{align}

Any consistent four-point function of scalar primary operators must obey the crossing relations:
\begin{align}
\begin{split}\label{crossing}
	G^{21}_{34}(x) &= G^{41}_{32}(1-x) \\
 &= |x|^{-4 \Delta_3 } G^{24}_{31}\left(\frac{1}{x}\right).
\end{split}
\end{align}
These relations follow from the invariance of \eqref{gen4pt} under exchange of any pair of indices, which we have already verified.  However as a check, we can verify \eqref{crossing} directly. By taking different limits of the four points we obtain
\begin{align}
	G^{41}_{32}(x) &= |x|^{-2 (\Delta_{14} - \Delta_2 - \Delta_3)} |1-x|^{- 2(\Delta_{12} - \Delta_3 - \Delta_4)}
	\exp\left[ {2 \tilde{\Delta}} A(x) \right] \label{G2}\\
	G^{24}_{31}(x) &= |x|^{-2(\Delta_{13} - \Delta_1 - \Delta_3)} |1-x|^{- 2(\Delta_{14} - \Delta_2 - \Delta_3)} \exp \left[ {2 \tilde{\Delta}A(x)}\right] . \label{G3}
\end{align}
 It is easy to see that \eqref{G1}, \eqref{G2} and \eqref{G3} indeed satisfy \eqref{crossing} (again making use of the identity \eqref{mysteriousident}).

\subsection{Primary operator spectrum}

As explained above, the  expansion of $G^{21}_{34}(x)$ around $x=0$ reveals the spectrum of dimensions of the primary operators that couple to the layering vertex operators.
The hypergeometric functions appearing in $A(x)$ are regular as $x \to 0$. As a result, the leading power comes from the term $ |x |^{-2(\Delta_{12} - \Delta_3 - \Delta_4)}$, where $\Delta_{12} = \Delta_{34} = {\frac{\lambda}{10}}(1 - \cos(\beta_3+\beta_4))$.
Therefore, the lightest operator with non-zero three point function with $e^{i \beta_3 N(z_3)} e^{i \beta_4 N(z_4)}$ and $e^{i \beta_1 N(z_1)} e^{i \beta_2 N(z_2)}$
 has dimension $\Delta_{12} = \Delta_{34}$, the dimension of the operator $e^{-i (\beta_3+\beta_4) N(z)} $. Furthermore, the three-point function coefficient is equal to 1.
Presumably, this operator is indeed $e^{- i(\beta_3+\beta_4) N(z)} = e^{ i(\beta_1+\beta_2) N(z)} $, although we cannot be certain as we do not have complete knowledge  of all its three-point function coefficients.
 
 The next  term in the 
 expansion of $G^{21}_{34}(x)$ comes from the $x^{1/3}{}_2F_1(2/3,1;4/3;x) = x^{1/3}(1 + \mathcal{O}(x))$ term. Since there are no other terms with the power $x^{1/3}$, there must be at least one primary operator with dimension $\Delta_{12} + 1/3$.
Similarly, expanding the exponential to quadratic order gives a term proportional to the square of the previous one, $x^{2/3}{}_2F_1(2/3,1;4/3;x)^2/2$. This indicates the existence of a primary with dimension $\Delta_{12} + 2/3$.

The question of whether there is a primary with  $\Delta_{12} + 3/3 = \Delta_{12} + 1$ is more subtle, because this power of $x$ also appears in the expansion of the $\Delta_{12}$ conformal block. To see that such an operator indeed exists, we could compute (the square of) its three-point function coefficient by subtracting the contribution from that of the $\Delta_{12}$ block, and see that the result is non-zero.

Let us now make this procedure systematic for the first few levels of operators.
As noted before, it appears there are operators with dimensions
\begin{align}\label{hp}
  \Delta^{(p,p')} = \Delta_{12} + \frac{p}{3} = \Delta_{34} + \frac{p}{3} = \frac{\lambda}{10}(1-\cos(\beta_1+\beta_2)) + \frac{p}{3}
\end{align}
for all non-negative integers $p$ that couple to $e^{i \beta_3 N(z_3)} e^{i \beta_4 N(z_4)}$ and $e^{i \beta_1 N(z_1)} e^{i \beta_2 N(z_2)}$.
We need to compare the  expansion of
\begin{align}\label{gtaylor}
    G^{2 1}_{3 4}(x) = x^{-\Delta_3} \bar{x}^{-\Delta_4} \sum_{m,n=0}^\infty a_{m n} x^{m/3} \bar{x}^{n/3},
\end{align}
to the conformal block expansion \eqref{confblexp}.
In the expansion we must allow the operators to have non-zero spin $s = \Delta - \bar{\Delta}$.
To accommodate the previous results we write
\begin{align}\label{confblexp2}
    G^{21}_{34}(x) = x^{-\Delta_3} \bar{x}^{-\Delta_4} \sum_{\substack{p,p',\\m,n=0}}^\infty C_{34}^{(p,p')} C_{12}^{(p,p')} ~ 
    \mathcal{F}_m^{(p)} \bar{\mathcal{F}}_n^{(p')} ~ x^{m+p/3} \bar{x}^{n+p'/3},
\end{align}
where we now sum over all non-negative integers $m,n,p,p'$. $F_m^{(p)}, \bar{F}_n^{(p')}$ denote the conformal block coefficients evaluated at \eqref{hp}.

By identifying the coefficients order by order in $x$ and $\bar{x}$ we can find the products of three-point coefficients $C_{34}^{(p,p')} C_{12}^{(p,p')}$. Every non-zero combination shows the existence of an operator with dimensions $H_p$ in the operator spectrum to which two operators fuse on to.

\begin{figure}[t]
    \centering
    \begin{subfigure}[b]{0.3\textwidth}
        \includegraphics[width=\textwidth]{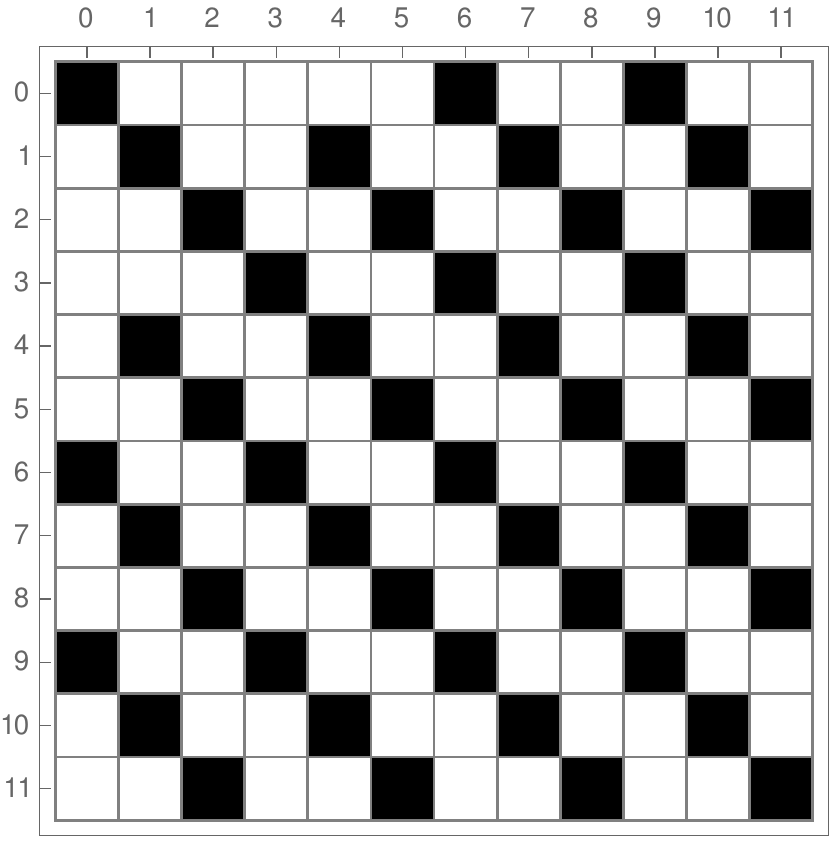}
        \caption{Generic $\beta_i$}
        \label{matrix_general}
    \end{subfigure}~
    \begin{subfigure}[b]{0.3\textwidth}
        \includegraphics[width=\textwidth]{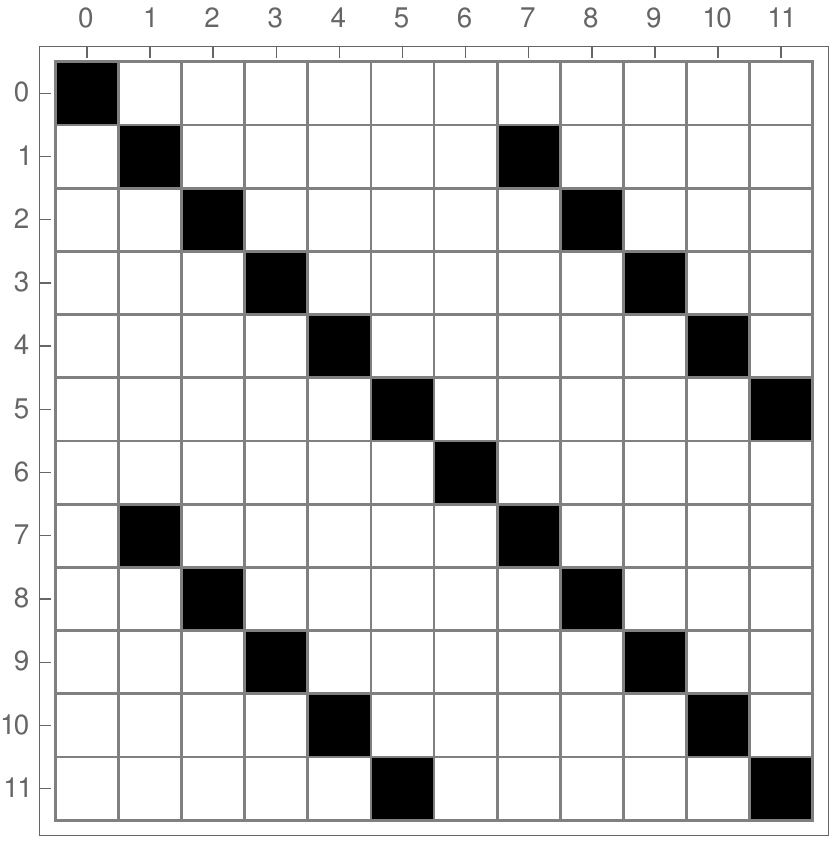}
        \caption{$\beta_1=\beta_2=\beta_3=\beta_4=\pi$}
        \label{matrix_pis}
    \end{subfigure}~
    \begin{subfigure}[b]{0.3\textwidth}
        \includegraphics[width=\textwidth]{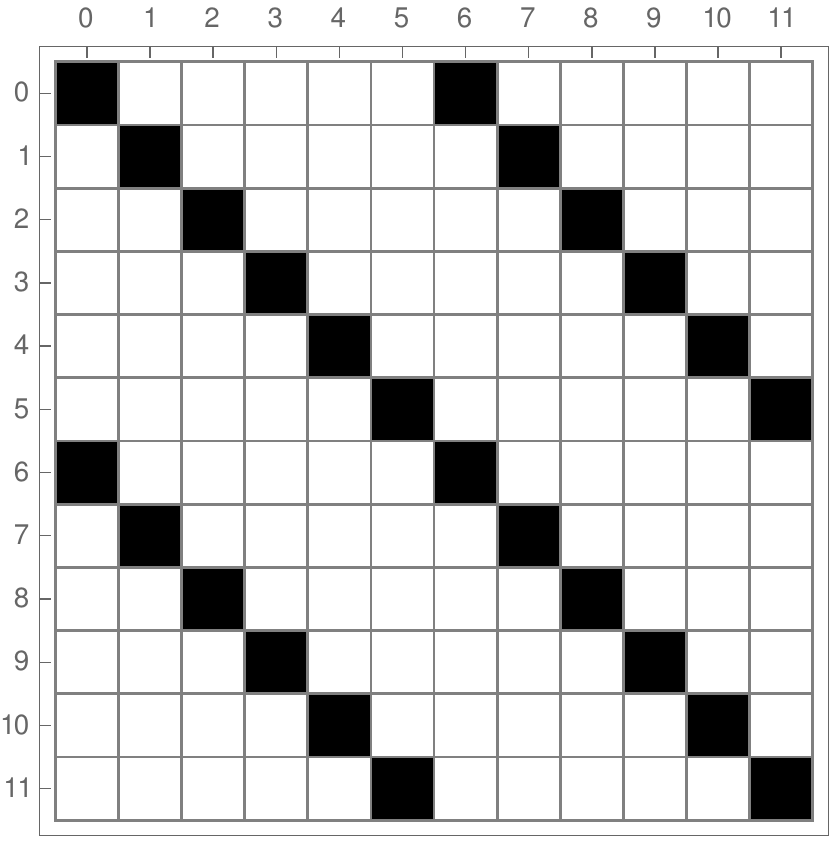}
        \caption{$\beta_1=\beta_2=\beta_3=\beta_4=\frac{\pi}{2}$}
        \label{fig:gull}
    \end{subfigure}
    \caption{The non-zero products $C_{34}^{(p, p')} C_{12}^{(p,p')}$ are shown for different choices of $\beta_i$.} 
    \label{matrix_figure}
\end{figure}

We use a code developed by Matt Headrick \cite{headrick} to generate the conformal block coefficients up to third order, which allows us to consider terms up to order $\mathcal{O}(x^{11/3})$.
The non-zero three point coefficients are marked in Fig.\ \ref{matrix_figure}.
We denote $\mu = \frac{8 \cdot 2^{1/3} \pi^2}{3 \sqrt{3} \Gamma(1/6)^2 \Gamma(4/3)^2}$.  Below, we give the first few three-point function coefficients from different blocks.  These grow very rapidly in complexity with increasing $p,p'$.
\begin{align}
    C_{34}^{(0,0)} C_{12}^{(0,0)} &=  1\\
    C_{34}^{(1,1)} C_{12}^{(1,1)} &=  -\frac{48}{5} \mu \lambda \sin\frac{\beta_1}{2} \sin\frac{\beta_2}{2} \sin\frac{\beta_3}{2} \sin\frac{\beta_1+\beta_2+\beta_3}{2}\\
    C_{34}^{(2,2)} C_{12}^{(2,2)} &= \frac{1}{2} \left(C_{34}^{(1,1)} C_{12}^{(1,1)}\right)^2 \\
    C_{34}^{(3,3)} C_{12}^{(3,3)} &= \frac{1}{3!} \left(C_{34}^{(1,1)} C_{12}^{(1,1)}\right)^3.
\end{align}
$C_{34}^{(4,4)} C_{12}^{(4,4)}$ and the following terms are very lengthy.
Off-diagonal terms of the form $C^{(n,n+3)}= C^{(n+3,n)}$:
\begin{align}
    C_{34}^{(0,3)} C_{12}^{(0,3)} &= 0 \\
    C_{34}^{(1,4)} C_{12}^{(1,4)} &= \frac{24 \mu \lambda^2}{5}
    \frac{\left(\cos \frac{\beta _3}{2} -\cos \left(\beta _1+\beta _2+\frac{3 \beta _3}{2}\right)\right)\sin \frac{\beta _1}{2} \sin \left(\frac{\beta _1-\beta _2}{2} \right) \sin \frac{\beta _2}{2} \sin \frac{\beta _3}{2}}{-10 - 3 \lambda (1 -  \cos \left(\beta _1+\beta _2\right) )} \\
    C_{34}^{(2,5)} C_{12}^{(2,5)} &= \frac{1152 \mu^2 \lambda^3}{25}
    \frac{
    (\cos(\beta_1+\beta_3) - \cos(\beta_2+\beta_3))
    \left( \sin\frac{\beta_1}{2} \sin\frac{\beta_2}{2} \sin\frac{\beta_3}{2} \sin\left(\frac{\beta_1 + \beta_2+\beta_3}{2}\right)
    \right)^2
    }{-20-3\lambda(1-\cos(\beta_1+\beta_2))} \\
    C_{34}^{(3,6)} C_{12}^{(3,6)} &=  -\frac{9216 \mu^3 \lambda^4}{125}
    \frac{
    (\cos(\beta_1+\beta_3) - \cos(\beta_2+\beta_3))
    \left( \sin\frac{\beta_1}{2} \sin\frac{\beta_2}{2} \sin\frac{\beta_3}{2} \sin\left(\frac{\beta_1 + \beta_2+\beta_3}{2}\right)
    \right)^3
    }{-10-\lambda(1-\cos(\beta_1+\beta_2))}
\end{align}
The first term of the form $C^{(n,n+6)}= C^{(n+6,n)}$ is
\begin{align}
\begin{split}
    C_{34}^{(0,6)} C_{12}^{(0,6)} = 
    &-\frac{\lambda^2}{200} [11+5(\cos{\beta_1}+\cos{\beta_2}-\cos(\beta_1+\beta_2)]\\
    &\times \sin\frac{\beta_1}{2} \sin\frac{\beta_2}{2} \sin\frac{\beta_3}{2}
    \Bigg[
    6\sin\frac{\beta_1+\beta_2-\beta_3}{2}+
    17\sin\frac{\beta_1+\beta_2+\beta_3}{2}\\
    &+ 5\sin\frac{3(\beta_1+\beta_2+\beta_3)}{2}-
    6\sin\frac{3(\beta_1+\beta)+\beta_3}{2}+
    5\sin \frac{\beta_1 + \beta_2 + 3 \beta_3}{2}
    \Bigg] \\
    &\times[
    25 + 14 \lambda
    + \cos(\beta_1+\beta_2) (25-\lambda(18-4\cos^2(\beta_1+\beta_2)))
    ]^{-1}.
\end{split}
\end{align}

We now analyze two special cases for which the three-point coefficients simplify considerably.
Consider first the case $\beta_1=\beta_2=\beta_3=\beta_4=\pi$.
We denote $C_{34}^{(p,p')} = C_{12}^{(p,p')} = C^{(p,p')}$. The first few diagonal terms $C^{(n,n)}$ are given for $0 \le n \le 6$ by
\begin{align}
    \left(C^{(n,n)}\right)^2 &= \frac{1}{n!} \left(C^{(1,1)}\right)^{2 n},
\end{align}
with
\begin{align}
    \left(C^{(1,1)}\right)^2 &= \frac{48}{5} \lambda \mu.
\end{align}
The term with $n=7$ is 
     \begin{align}
    \left(C^{(7,7)}\right)^2 &= \frac{768 \mu \lambda^3}{19140625}
    \left(
        \frac{125}{(7-15 \lambda)^2} + 37158912 \mu^6 \lambda^4
    \right).
\end{align}
The first few off-diagonal terms are of the form $C^{(n,n+6)}= C^{(n+6,n)}$
\begin{align}
\begin{split}
    \left(C^{(6,0)}\right)^2 &= 0\\
    \left(C^{(7,1)}\right)^2 &= -\frac{192}{875} \frac{\lambda^2 \mu}{15 \lambda - 7}\\
    \left(C^{(8,2)}\right)^2 &= \frac{288}{4375} \frac{161\lambda - 50}{21 \lambda + 2} \lambda^2 \mu^2\\
    \left(C^{(9,3)}\right)^2 &= \frac{1536}{21875} \frac{59\lambda - 25}{\lambda + 1} \lambda^3 \mu^3.
\end{split}
\end{align}

Now consider the case $\beta_1=\beta_2=\beta_3=\beta_4=\frac{\pi}{2}$.
The diagonal terms for $0 \le n \le 5$ are
\begin{align}
    \left(C^{(n,n)}\right)^2 &= \frac{1}{n!} \left(C^{(1,1)}\right)^{2 n}
\end{align}
with
\begin{align}
    \left(C^{(1,1)}\right)^2 &= -\frac{12}{5} \lambda \mu.
\end{align}
The $n=6$ term is
\begin{align}
    \left(C^{(6,6)}\right)^2 &= \frac{\lambda^2}{1250000} (125 + 331776 \mu^6 \lambda^4)
\end{align}
The first few off-diagonal terms $C^{n,n+6}$ are
\begin{align}
\begin{split}
    \left(C^{(6,0)}\right)^2  &= -\frac{\lambda}{100}\\
    \left(C^{(7,1)}\right)^2 &= 
    \frac{3 \lambda^2 \mu}{1750}
    \frac{2268\lambda^2+5835\lambda-1450}
    {162\lambda^2+390\lambda-175}
    \\
    \left(C^{(8,2)}\right)^2 &= 
    -\frac{9 \lambda^2 \mu^2}{4375}
    \frac{1134 \lambda^3 + 5835 \lambda^2 + 2350 \lambda + 1250}
    {81\lambda^2 + 390 \lambda + 25}
    \\
    \left(C^{(9,3)}\right)^2 &=
    \frac{36 \lambda^3 \mu^3}{21875}
    \frac{252\lambda^3+1945\lambda^2+2050\lambda+1250}
    {18\lambda^2+130\lambda+75}.
\end{split}
\end{align}

\subsection{Interpretation}

We leave the physical interpretation of these new primaries to future work. A hint is provided by \cite{Simmons_2009}, which considers the $O(n)$ model as $n \to 0$.  There the four-point function is essentially the log of the one considered here for $\beta_i = \pi$, and only a finite number of primaries appear in the fusion products.  The primary corresponding $(p=1, p'=1)$ in our notation has dimension $\left(\frac{1}{3},\frac{1}{3}\right)$ when $\beta_i = \pi$, and is identified as the leading order energy density operator of the $O(n)$ model (in the limit $n \to 0$).

One important caveat to our results in this section is that the three-point function coefficients obtained from the conformal block expansion do not entirely determine the spectrum of primaries.  Clearly, there could be primaries in the theory with vanishing three-point coefficients with the layering vertex operators, and these would be invisible to us.  A more subtle issue can also arise in the other direction if there are multiple operators with the same conformal dimensions that couple to the vertex operators.  In that case one can only determine the sum of the squares of the three-point coefficients.  Since these squared coefficients can evidently be negative, there could be cancellations.  Therefore it is logically possible we are missing some operators in the theory that couple to the vertex operators, as there could be multiple degenerate primaries that couple with three-point coefficients with squares that sum to zero.

\subsection{Null descendant states}

Some of the three-point function coefficients we have calculated diverge at special values of $\lambda, \beta_i$.  For instance, with all $\beta_i=\pi$ we have
\begin{align}
\left(C^{(7,1)}\right)^2 &= -\frac{192}{875} \frac{\lambda^2 \mu}{15 \lambda - 7}.
\end{align}
This coefficient diverges when $\lambda = 7/15$, or $c=14/15$.  One expects CFTs with $c<1$ to contain null descendants of primaries with certain conformal dimensions.  Indeed, at $c = 14/15$ the Kac determinant vanishes at second level for a primary with dimension $\Delta = 1/3$ ($h_{2,1}$ in standard notation, see for instance \cite{DiFrancesco:639405}).  Vanishing at second level means a state with dimension $1/3 + 2 = 7/3$ should become null.  When this happens the coefficient $C$ will diverge, because the norm of the state appears in the denominator.  The dimension of the operator corresponding to $C^{(7,1)}$ is indeed $\Delta^{(7,1)} = (\lambda/10)(1-\cos (\beta_1+\beta_2))+ p/3 = 7/3$ for $\beta_i = \pi$ and $p=7$, as expected from this argument.\footnote{We thank Alex Maloney and Liam Fitzpatrick for discussions on this point.}

\section{Outlook}

Our results for the correlation functions raise many interesting questions.  First, it is possible that we can extend our techniques to compute $n$-point correlation functions for arbitrary $n$.  This would provide  new results for the winding probabilities of Brownian loops.  The spectrum of new primary operators we discovered needs investigation, as we do not know how to identify these operators either in terms of a previously known CFT, or in terms of the BLS.  

Another interesting direction is to generalize the random variables assigned to the loops.  Here we considered the layering operator and assigned a random $\pm 1$ to each loop.  In ongoing work to appear soon, two of us (Foit and Kleban) have considered more general distributions of random weights.  This gives rise to infinite class of new conformally invariant systems for which we can compute exact four-point functions that depend on additional continuous parameters  characterizing the distribution of weights.

\begin{acknowledgments}
We would like to thank Liam Fitzpatrick, Ben Freivogel, Gaston Giribet, Matthew Headrick,  Alex Maloney, and Massimo Porrati for useful discussions. The work of V.\ F.\ is supported by the James Arthur Graduate Award.  The work of M.\ K.\ is supported by the NSF through the grant PHY-1820814. 
\end{acknowledgments}

\appendix
\section{}\label{appendix}

In this appendix we derive the results in Sec.\ \ref{sec23full}
by taking limits of our results from the upper half-plane.

We consider the two-point functions first and apply charge conservation
$\Delta_{12}=0$. Taking
\begin{equation} \label{eq 3.1}
    \zeta_1(t)=x_1 +i (y_1 + t), \quad
    \zeta_2(t)=x_2 +i( y_2 + t)
\end{equation}
with $t \geq 0$ and $z_1, z_2$ as before,
we can express the two-point function in the complex plane as the two-point function in the upper half-plane with both points far away from the boundary
\begin{equation}\label{planelimit}
 \Braket{e^{i \beta_1 N(z_1)} e^{i \beta_2 N(z_2)}}_{\mathbb C}  
  = \lim_{t \to \infty}\Braket{e^{i \beta_1 N(\zeta_1(t))} e^{i \beta_2 N(\zeta_2(t))}}_{\mathbb{H}}.
\end{equation}
To see this, note that 
\begin{align}
\begin{split}
 \Braket{e^{i \beta_1 N(\zeta_1(t))} e^{i \beta_2 N(\zeta_2(t))}}_{\mathbb{H}}
 &= \Braket{e^{i \beta_1 N(z_1)} e^{i \beta_2 N(z_2)}}_{\mathbb{H}_t} \\
 &= e^{-\lambda \alpha_{\mathbb{H}_t}(z_1|z_2)(1-\cos\beta_1)}
 e^{-\lambda \alpha_{\mathbb{H}_t}(z_2|z_1)(1-\cos\beta_2)},
\end{split}
\end{align}
where $\mathbb{H}_t$ is the half-plane $\{(x,y): y \geq -t\}$.
The weights $\alpha_{\mathbb{H}_t}(z_j|z_k)$,  for unequal $j,k=1,2$, are increasing in $t$ and bounded above by $\alpha(z_j|z_k)$,
 which is finite by thinness of the Brownian loop measure
$\mu^{\text{loop}}$ \cite{Nacu_2011}; this implies that
they have a finite limit as $t \to \infty$.
Moreover,
\begin{align}
    \alpha(z_j|z_k) - \alpha_{\mathbb{H}_t}(z_j|z_k) \leq \mu^{\text{loop}}(\gamma: \operatorname{diam}(\gamma) \geq t, \gamma \text{ intersect } \overline{z_1 z_2}) \to 0,
\end{align}
 where $\overline{z_1 z_2}$ is the segment connecting
$z_1$ and $z_2$, as $t \to \infty$, again by thinness. This shows that $\lim_{t \to \infty} \alpha_{\mathbb{H}_t}(z_i|z_j) = \alpha(z_i|z_j)$ and proves \eqref{planelimit}.
Notice also that
\begin{align}
 \lim_{t \to \infty} \frac{\zeta_1(t)-\overline \zeta_1(t)}
 {\zeta_1(t)-\overline \zeta_2(t)} = \lim_{t \to \infty}
 \frac{2 i (y_1 + t)}{(x_1-x_2)+ i(y_1 + y_2 + 2t)}=1, \\
 \lim_{t \to \infty} \frac{\zeta_2(t)-\overline \zeta_2(t)}
 {\zeta_1(t)-\overline \zeta_2(t)} = \lim_{t \to \infty}
 \frac{2 i (y_2+t)}{(x_1-x_2)+ i(y_1 + y_2 + 2t)}=1.
\end{align}
It follows that
\begin{align}
 \lim_{t \to \infty} \eta=0,\quad
 \lim_{t \to \infty} \sigma=0, \quad
 \lim_{t \to \infty} \frac{\sigma}{\eta}=1
\end{align}
and
\begin{align}
    \lim_{\eta \to 0} \eta {}_3 F_{2}\left(1,1,\frac{4}{3};2,\frac{5}{3}; \eta \right)
    =\lim_{\eta \to 0} \eta {}_2 F_{1}\left(1,\frac 2 3;\frac{4}{3}, \eta\right) =0.
\end{align}
From the equality of \eqref{eq 1} and \eqref{eq 2} we have that
\begin{eqnarray}
 \lim_{t \to \infty} (1-\sigma)
 {}_3 F_{2}\left(1,1,\frac{4}{3};2,\frac{5}{3}; 1-\sigma \right)
 =\frac{2 \pi}{\sqrt 3}.
\end{eqnarray}
With charge conservation $\beta_1+\beta_2 = 2 \pi \mathbb{Z}$, we  have from \eqref{eq 3} that
\begin{align}
\begin{split}\label{eq 3.2}
 & \lim_{t \to \infty} \lim_{\delta \to 0}
   \left(2\delta e^{-5 \overline \alpha}\right)^{-2(\Delta_1+\Delta_2)
 }\Braket{e^{i \beta_1 N(z_1)} e^{i \beta_2 N(z_2)}}_{\mathbb{H}} \\
   =& \lim_{t \to \infty}
  |z_1(t)-z_2(t)|^{-2(\Delta_1+\Delta_2)}
     |z_1(t)-\overline z_2(t)|^{2(\Delta_1+\Delta_2)}\\
  &   \quad |z_1(t)-\overline z_1(t)|^{-2\Delta_1}  |z_2(t)-\overline z_2(t)|^{-2\Delta_2}  \\
  &   \quad  \times \exp \left[ -(\Delta_1+\Delta_2)(1-\sigma) {}_3 F_{2}\left(1,1,\frac{4}{3};2,\frac{5}{3};  1-\sigma \right) \right] \\
  =& \left(e^{\frac{\pi}{\sqrt{3}}}\right)^{-4\Delta_1}
   |z_1-z_2|^{-4\Delta_1},
\end{split}
\end{align}
where we used the fact that $\Delta_1 = \Delta_2$.  This gives an explicit expression of the constant appearing
in the two-point function for the plane in \cite{camia2016brownian};
with the normalization used in \cite{camia2016brownian} (see the Summary and Results section, below (2.3))
the constant $C_2$ defined there is 
\begin{equation}
    C_2=\left(2e^{-\frac{\pi}{\sqrt{3}}-5  \bar\alpha } \right)^{2 (\Delta_1+\Delta_{2})}=\left(2e^{-\frac{\pi}{\sqrt{3}}-5  \bar\alpha } \right)^{4 \Delta_1}.
\end{equation}
Absorbing the constants $\left(2 e^{-\frac{\pi}{\sqrt{3}}-5  \bar\alpha }\right)^{2 \Delta_j}$ into the definition of ${\mathcal O}_{\beta_j}$ gives the canonically normalized two-point function in the plane
\begin{align}\label{2pt}
    \Braket{{\mathcal O}_{\beta_1}(z_1) {\mathcal O}_{\beta_2}(z_2)}_\mathbb{C} = \lim_{\delta \to 0}
    \frac{ \Braket{e^{i \beta_1 N(z_1)} e^{i \beta_2 N(z_2)}}_\mathbb{C} }
    {\left(2 \delta e^{-\frac{\pi}{\sqrt{3}}-5  \bar\alpha }\right)^{2 (\Delta_1+
    \Delta_2)}}
    =|z_1-z_2|^{-4 \Delta_1}.
\end{align}

It turns out we can also compute the three-point functions in the full plane using only the weights for the Brownian loop measure encircling one and two points in the upper half-plane. Given
$z_i, z_j, z_k \in \mathbb C$, for distinct $i, j, k
\in \{1,2,3\}$ and $\delta \leq \min_{i j} |z_i-z_j|$,
we have the six relations
\begin{align}
\begin{split}
\alpha_{\delta, \mathbb{H}}(z_i)&=
\alpha_{\mathbb{H}}(z_i|z_j,z_k)+
\alpha_{ \mathbb{H}}(z_i, z_j|z_k)
+\alpha_{ \mathbb{H}}(z_i, z_k|z_j)+\alpha_{ \mathbb{H}}(z_1, z_2,z_3)\\
 \alpha_{ \mathbb{H}}(z_i, z_j) &=
 \alpha_{ \mathbb{H}}(z_i, z_j|z_k)+\alpha_{ \mathbb{H}}(z_1, z_2,z_3)
\end{split}
\end{align}
which give
\begin{align}
\begin{split}\label{4.10}
\alpha_{ \mathbb{H}}(z_i, z_j|z_k)&=
\alpha_{ \mathbb{H}}(z_i, z_j)-\alpha_{ \mathbb{H}}(z_1, z_2,z_3)\\
\alpha_{\mathbb{H}}(z_i|z_j,z_k)&=
\alpha_{\mathbb{H}}(z_i)-\alpha_{ \mathbb{H}}(z_i, z_j)
-\alpha_{ \mathbb{H}}(z_i, z_k)+\alpha_{ \mathbb{H}}(z_1, z_2,z_3).
\end{split}
\end{align}
It follows from \eqref{npoint} that
\begin{align}
\begin{split}\label{6.11}
    &\Braket{e^{i \beta_1 N(z_1)} e^{i \beta_2 N(z_2)}e^{i \beta_3 N(z_3)}}_\mathbb{H}\\
    =& \exp\Bigg[  - \lambda \Bigg( \sum_{j=1}^3 (1-\cos\beta_j)  \alpha_{\mathbb{H}}(z_j|z_i, z_k)
    + \sum_{\substack{j,k=1 \\ j < k}}^3 (1-\cos(\beta_j+\beta_k)) \alpha_{ \mathbb{H}}(z_j, z_k|z_i) \\
    & \quad  + (1-\cos(\beta_1+\beta_2+\beta_3))\alpha_{ \mathbb{H}}(z_1, z_2,z_3) \Bigg) \Bigg] \\
    =& \exp\Bigg[ - \lambda \Bigg( \sum_{j=1}^3 (1-\cos\beta_j) \alpha_{\delta, \mathbb{H}}(z_j) \\
    &  \quad + \left( \sum_{\substack{j,k=1\\j<k}}^3(-(1-\cos\beta_j)-(1-\cos\beta_k) + (1-\cos(\beta_j+\beta_k)) \right) \alpha_{\mathbb{H}}(z_j, z_k) \\
  & \quad + \left( \sum_{j=1}^3 (1-\cos\beta_j)-\sum_{\substack{j,k=1\\j<k}}^3 (1-\cos(\beta_j+\beta_k))+1-\cos(\beta_1+\beta_2+\beta_3) \right) \alpha_{ \mathbb{H}}(z_1,z_2,z_3) \Bigg) \Bigg]
\end{split}
\end{align}
With charge conservation $\beta_1+\beta_2+\beta_3=2 \pi \mathbb Z$ it follows that $\cos(\beta_i+\beta_j)=\cos(\beta_k)$, and hence the coefficient of $\alpha_{ \mathbb{H}}(z_1, z_2,z_3)$ is identically
zero. This gives the three-point function 
\begin{align}
\begin{split}
    &\Braket{e^{i \beta_1 N(z_1)} e^{i \beta_2 N(z_2)}e^{i \beta_3 N(z_3)}}_\mathbb{H}\\
    =& \left(2\delta e^{-5 \overline \alpha} \right)^{2 (\Delta_1 + \Delta_2 + \Delta_3)}
    \prod_{i=1}^3 |z_i-\overline z_i|^{-2\Delta_i}
    \prod_{\substack{j,k=1\\j<k}}^3 \Bigg( |z_j-z_k|^{-2(\Delta_j+\Delta_k-\Delta_{j k})}
     |z_j-\overline z_k|^{2(\Delta_j+\Delta_k-\Delta_{j k})} \\
     & \quad \quad
 \times \exp\left[-\lambda (\Delta_j+\Delta_k- \Delta_{j k})(1-\sigma_{j k}) {}_3 F_{2}\left(1,1,\frac{4}{3};2,\frac{5}{3};  1-\sigma_{j k} \right) \right] \Bigg),
\end{split}
\end{align}
where $\sigma_{j k} = \frac{|z_j - z_k|^2}{|z_j - \overline z_k|^2}$. 
Again, using charge conservation, and inserting the canonical normalization factors found for the two point functions, we derive the three-point function in the full plane by taking the limit (analogous to \eqref{planelimit})
\begin{align}
\begin{split}\label{eq 3.5}
& \Braket{{\mathcal O}_{\beta_1}(z_1) {\mathcal O}_{\beta_2}(z_2) {\mathcal O}_{\beta_3}(z_3)}_\mathbb{C}\\
 =& \lim_{t \to \infty} \lim_{\delta \to 0} \left(2\delta e^{-5 \overline \alpha -\frac{\pi}{\sqrt 3}}\right)^{-2 \sum_{j=1}^3\Delta_j }
\Braket{e^{i \beta_1 N(z_1)} e^{i \beta_2 N(z_2)}e^{i \beta_3 N(z_3)}}_\mathbb{H}\\
  =& \lim_{t \to \infty}  \left( e^{
-\frac{\pi}{\sqrt 3}}
\right)^{-2 \sum_{j=1}^3\Delta_j }
 \prod_{i=1}^3 |z_i(t)-\overline z_i(t)|^{-2\Delta_i} \\
  &  \quad \prod_{\substack{j,k=1 \\ j<k}}^3 \Bigg( |z_j(t)-z_k(t)|^{-2(\Delta_j+\Delta_k-\Delta_{j k})}
     |z_j(t)-\overline z_k(t)|^{2(\Delta_j+\Delta_k-\Delta_{j k})} \\
  & \quad \times \exp \left[ -(\Delta_j+\Delta_k- \Delta_{j k})(1-\sigma) {}_3 F_{2}\left(1,1,\frac{4}{3};2,\frac{5}{3};  1-\sigma \right)
  \right] \Bigg) \\
  =&
   |z_1-z_2|^{-2(\Delta_1+\Delta_2-\Delta_{1 2})}
     |z_1- z_3|^{-2(\Delta_1+\Delta_3-\Delta_{1 3})}
     |z_2- z_3|^{-2(\Delta_2+\Delta_3-\Delta_{2 3})} \\
  =&
   |z_1-z_2|^{-2(\Delta_1+\Delta_2-\Delta_{3})}
     |z_1- z_3|^{-2(\Delta_1+\Delta_3-\Delta_{2})}
     |z_2- z_3|^{-2(\Delta_2+\Delta_3-\Delta_{1})},
\end{split}
\end{align}
which has the correct $z_i$ dependence for a three-point function.
Surprisingly, the overall coefficient of the
three-point function is simply $1$, and does not depend on the $\beta_i$ (this was also the case for the three-point functions in the disk model of \cite{Freivogel:2009rf}).

From this result we obtain an explicit expression for the constant appearing
in the three-point function in \cite{camia2016brownian};
with the normalization used there (see the Summary and Results section, below (2.3))
\begin{align}
    C_3 =\left(2e^{-\frac{\pi}{\sqrt{3}}-5  \bar\alpha}\right)^{2(\Delta_1+\Delta_{2}+\Delta_{3})}.
\end{align}

\bibliography{bibliography}

\begin{thebibliography}{13}%
\makeatletter
\providecommand \@ifxundefined [1]{%
 \@ifx{#1\undefined}
}%
\providecommand \@ifnum [1]{%
 \ifnum #1\expandafter \@firstoftwo
 \else \expandafter \@secondoftwo
 \fi
}%
\providecommand \@ifx [1]{%
 \ifx #1\expandafter \@firstoftwo
 \else \expandafter \@secondoftwo
 \fi
}%
\providecommand \natexlab [1]{#1}%
\providecommand \enquote  [1]{``#1''}%
\providecommand \bibnamefont  [1]{#1}%
\providecommand \bibfnamefont [1]{#1}%
\providecommand \citenamefont [1]{#1}%
\providecommand \href@noop [0]{\@secondoftwo}%
\providecommand \href [0]{\begingroup \@sanitize@url \@href}%
\providecommand \@href[1]{\@@startlink{#1}\@@href}%
\providecommand \@@href[1]{\endgroup#1\@@endlink}%
\providecommand \@sanitize@url [0]{\catcode `\\12\catcode `\$12\catcode
  `\&12\catcode `\#12\catcode `\^12\catcode `\_12\catcode `\%12\relax}%
\providecommand \@@startlink[1]{}%
\providecommand \@@endlink[0]{}%
\providecommand \url  [0]{\begingroup\@sanitize@url \@url }%
\providecommand \@url [1]{\endgroup\@href {#1}{\urlprefix }}%
\providecommand \urlprefix  [0]{URL }%
\providecommand \Eprint [0]{\href }%
\providecommand \doibase [0]{http://dx.doi.org/}%
\providecommand \selectlanguage [0]{\@gobble}%
\providecommand \bibinfo  [0]{\@secondoftwo}%
\providecommand \bibfield  [0]{\@secondoftwo}%
\providecommand \translation [1]{[#1]}%
\providecommand \BibitemOpen [0]{}%
\providecommand \bibitemStop [0]{}%
\providecommand \bibitemNoStop [0]{.\EOS\space}%
\providecommand \EOS [0]{\spacefactor3000\relax}%
\providecommand \BibitemShut  [1]{\csname bibitem#1\endcsname}%
\let\auto@bib@innerbib\@empty
\bibitem [{\citenamefont {Lawler}\ and\ \citenamefont
  {Werner}(2004)}]{lawler2003brownian}%
  \BibitemOpen
  \bibfield  {author} {\bibinfo {author} {\bibfnamefont {G.~F.}\ \bibnamefont
  {Lawler}}\ and\ \bibinfo {author} {\bibfnamefont {W.}~\bibnamefont
  {Werner}},\ }\href {\doibase 10.1007/s00440-003-0319-6} {\bibfield  {journal}
  {\bibinfo  {journal} {Probability Theory and Related Fields}\ }\textbf
  {\bibinfo {volume} {128}},\ \bibinfo {pages} {565} (\bibinfo {year}
  {2004})}\BibitemShut {NoStop}%
\bibitem [{\citenamefont {Camia}\ \emph {et~al.}(2016)\citenamefont {Camia},
  \citenamefont {Gandolfi},\ and\ \citenamefont {Kleban}}]{camia2016brownian}%
  \BibitemOpen
  \bibfield  {author} {\bibinfo {author} {\bibfnamefont {F.}~\bibnamefont
  {Camia}}, \bibinfo {author} {\bibfnamefont {A.}~\bibnamefont {Gandolfi}}, \
  and\ \bibinfo {author} {\bibfnamefont {M.}~\bibnamefont {Kleban}},\ }\href
  {\doibase 10.1016/j.nuclphysb.2015.11.022} {\bibfield  {journal} {\bibinfo
  {journal} {Nucl. Phys.}\ }\textbf {\bibinfo {volume} {B902}},\ \bibinfo
  {pages} {483} (\bibinfo {year} {2016})},\ \Eprint
  {http://arxiv.org/abs/1501.05945} {arXiv:1501.05945 [math-ph]} \BibitemShut
  {NoStop}%
\bibitem [{\citenamefont {Han}\ \emph {et~al.}(2017)\citenamefont {Han},
  \citenamefont {Wang},\ and\ \citenamefont {Zinsmeister}}]{han2017brownian}%
  \BibitemOpen
  \bibfield  {author} {\bibinfo {author} {\bibfnamefont {Y.}~\bibnamefont
  {Han}}, \bibinfo {author} {\bibfnamefont {Y.}~\bibnamefont {Wang}}, \ and\
  \bibinfo {author} {\bibfnamefont {M.}~\bibnamefont {Zinsmeister}},\
  }\href@noop {} {\enquote {\bibinfo {title} {On the brownian loop measure},}\
  } (\bibinfo {year} {2017}),\ \Eprint {http://arxiv.org/abs/1707.00965}
  {arXiv:1707.00965 [math-ph]} \BibitemShut {NoStop}%
\bibitem [{\citenamefont {Gamsa}\ and\ \citenamefont
  {Cardy}(2006)}]{Gamsa_2006}%
  \BibitemOpen
  \bibfield  {author} {\bibinfo {author} {\bibfnamefont {A.}~\bibnamefont
  {Gamsa}}\ and\ \bibinfo {author} {\bibfnamefont {J.}~\bibnamefont {Cardy}},\
  }\href {\doibase 10.1088/0305-4470/39/41/s12} {\bibfield  {journal} {\bibinfo
   {journal} {Journal of Physics A: Mathematical and General}\ }\textbf
  {\bibinfo {volume} {39}},\ \bibinfo {pages} {12983–13003} (\bibinfo {year}
  {2006})}\BibitemShut {NoStop}%
\bibitem [{\citenamefont {Werner}(2008)}]{werner2005conformally}%
  \BibitemOpen
  \bibfield  {author} {\bibinfo {author} {\bibfnamefont {W.}~\bibnamefont
  {Werner}},\ }\href {\doibase 10.1090/S0894-0347-07-00557-7} {\bibfield
  {journal} {\bibinfo  {journal} {J. Amer. Math. Soc.}\ }\textbf {\bibinfo
  {volume} {21}},\ \bibinfo {pages} {137} (\bibinfo {year} {2008})}\BibitemShut
  {NoStop}%
\bibitem [{\citenamefont {Beliaev}\ and\ \citenamefont
  {Viklund}(2013)}]{Beliaev13}%
  \BibitemOpen
  \bibfield  {author} {\bibinfo {author} {\bibfnamefont {D.}~\bibnamefont
  {Beliaev}}\ and\ \bibinfo {author} {\bibfnamefont {F.~J.}\ \bibnamefont
  {Viklund}},\ }\href@noop {} {\bibfield  {journal} {\bibinfo  {journal}
  {Communications in Mathematical Physics}\ }\textbf {\bibinfo {volume}
  {320}},\ \bibinfo {pages} {379–394} (\bibinfo {year} {2013})}\BibitemShut
  {NoStop}%
\bibitem [{\citenamefont {Freivogel}\ and\ \citenamefont
  {Kleban}(2009)}]{Freivogel:2009rf}%
  \BibitemOpen
  \bibfield  {author} {\bibinfo {author} {\bibfnamefont {B.}~\bibnamefont
  {Freivogel}}\ and\ \bibinfo {author} {\bibfnamefont {M.}~\bibnamefont
  {Kleban}},\ }\href {\doibase 10.1088/1126-6708/2009/12/019} {\bibfield
  {journal} {\bibinfo  {journal} {JHEP}\ }\textbf {\bibinfo {volume} {12}},\
  \bibinfo {pages} {019} (\bibinfo {year} {2009})},\ \Eprint
  {http://arxiv.org/abs/0903.2048} {arXiv:0903.2048 [hep-th]} \BibitemShut
  {NoStop}%
\bibitem [{\citenamefont {Freivogel}\ \emph {et~al.}(2009)\citenamefont
  {Freivogel}, \citenamefont {Kleban}, \citenamefont {Nicolis},\ and\
  \citenamefont {Sigurdson}}]{Freivogel:2009it}%
  \BibitemOpen
  \bibfield  {author} {\bibinfo {author} {\bibfnamefont {B.}~\bibnamefont
  {Freivogel}}, \bibinfo {author} {\bibfnamefont {M.}~\bibnamefont {Kleban}},
  \bibinfo {author} {\bibfnamefont {A.}~\bibnamefont {Nicolis}}, \ and\
  \bibinfo {author} {\bibfnamefont {K.}~\bibnamefont {Sigurdson}},\ }\href
  {\doibase 10.1088/1475-7516/2009/08/036} {\bibfield  {journal} {\bibinfo
  {journal} {JCAP}\ }\textbf {\bibinfo {volume} {0908}},\ \bibinfo {pages}
  {036} (\bibinfo {year} {2009})},\ \Eprint {http://arxiv.org/abs/0901.0007}
  {arXiv:0901.0007 [hep-th]} \BibitemShut {NoStop}%
\bibitem [{\citenamefont {Camia}\ \emph {et~al.}(2019)\citenamefont {Camia},
  \citenamefont {Gandolfi}, \citenamefont {Peccati},\ and\ \citenamefont
  {Reddy}}]{camia2019brownian}%
  \BibitemOpen
  \bibfield  {author} {\bibinfo {author} {\bibfnamefont {F.}~\bibnamefont
  {Camia}}, \bibinfo {author} {\bibfnamefont {A.}~\bibnamefont {Gandolfi}},
  \bibinfo {author} {\bibfnamefont {G.}~\bibnamefont {Peccati}}, \ and\
  \bibinfo {author} {\bibfnamefont {T.~R.}\ \bibnamefont {Reddy}},\ }\href@noop
  {} {\enquote {\bibinfo {title} {Brownian loops, layering fields and imaginary
  gaussian multiplicative chaos},}\ } (\bibinfo {year} {2019}),\ \Eprint
  {http://arxiv.org/abs/1908.05881} {arXiv:1908.05881 [math.PR]} \BibitemShut
  {NoStop}%
\bibitem [{\citenamefont {Nacu}\ and\ \citenamefont
  {Werner}(2011)}]{Nacu_2011}%
  \BibitemOpen
  \bibfield  {author} {\bibinfo {author} {\bibfnamefont {{\c{S}}.}~\bibnamefont
  {Nacu}}\ and\ \bibinfo {author} {\bibfnamefont {W.}~\bibnamefont {Werner}},\
  }\href {\doibase 10.1112/jlms/jdq094} {\bibfield  {journal} {\bibinfo
  {journal} {Journal of the London Mathematical Society}\ }\textbf {\bibinfo
  {volume} {83}},\ \bibinfo {pages} {789–809} (\bibinfo {year}
  {2011})}\BibitemShut {NoStop}%
\bibitem [{\citenamefont {Di~Francesco}\ \emph {et~al.}(1997)\citenamefont
  {Di~Francesco}, \citenamefont {Mathieu},\ and\ \citenamefont
  {Sénéchal}}]{DiFrancesco:639405}%
  \BibitemOpen
  \bibfield  {author} {\bibinfo {author} {\bibfnamefont {P.}~\bibnamefont
  {Di~Francesco}}, \bibinfo {author} {\bibfnamefont {P.}~\bibnamefont
  {Mathieu}}, \ and\ \bibinfo {author} {\bibfnamefont {D.}~\bibnamefont
  {Sénéchal}},\ }\href {\doibase 10.1007/978-1-4612-2256-9} {\emph {\bibinfo
  {title} {{Conformal field theory}}}},\ Graduate texts in contemporary
  physics\ (\bibinfo  {publisher} {Springer},\ \bibinfo {address} {New York,
  NY},\ \bibinfo {year} {1997})\BibitemShut {NoStop}%
\bibitem [{\citenamefont {Headrick}()}]{headrick}%
  \BibitemOpen
  \bibfield  {author} {\bibinfo {author} {\bibfnamefont {M.}~\bibnamefont
  {Headrick}},\ }\href@noop {} {\enquote {\bibinfo {title} {Mathematica
  packages},}\ }\bibinfo {howpublished}
  {\url{http://people.brandeis.edu/~headrick/Mathematica/index.html}},\
  \bibinfo {note} {accessed: 2019-08-05}\BibitemShut {NoStop}%
\bibitem [{\citenamefont {Simmons}\ and\ \citenamefont
  {Cardy}(2009)}]{Simmons_2009}%
  \BibitemOpen
  \bibfield  {author} {\bibinfo {author} {\bibfnamefont {J.~J.~H.}\
  \bibnamefont {Simmons}}\ and\ \bibinfo {author} {\bibfnamefont
  {J.}~\bibnamefont {Cardy}},\ }\href {\doibase 10.1088/1751-8113/42/23/235001}
  {\bibfield  {journal} {\bibinfo  {journal} {Journal of Physics A:
  Mathematical and Theoretical}\ }\textbf {\bibinfo {volume} {42}},\ \bibinfo
  {pages} {235001} (\bibinfo {year} {2009})}\BibitemShut {NoStop}%
\end{thebibliography}%
	
\end{document}